\documentclass[aps,prb,twocolumn,english,superscriptaddress,citeautoscript,preprintnumbers,amsmath,amssymb,floatfix,footinbib]{revtex4-2}
\usepackage{float}
\usepackage[breaklinks=true,colorlinks,citecolor=blue,linkcolor=blue,urlcolor=blue]{hyperref}
\usepackage{amsmath}
\usepackage{graphicx}
\usepackage{dcolumn}
\usepackage{float}
\usepackage[utf8]{inputenc}
\usepackage{color}
\usepackage{placeins}

\begin{document}
	
    \title{Orbital Selective Dirac-like States in EuAgAs Revealed by Polarization Dependent ARPES and DFT}
	\author{Mohit Mudgal} \thanks{These authors contributed equally to this work.}
	\affiliation{Research Institute for Synchrotron Radiation Science (HISOR), Hiroshima University, Higashi-Hiroshima 739-0046, Japan}
    \author{Suman Nandi} \thanks{These authors contributed equally to this work.}
	\affiliation{Research Institute for Synchrotron Radiation Science (HISOR), Hiroshima University, Higashi-Hiroshima 739-0046, Japan}
 	\author{Mohamed El Gazzah}
	\affiliation{Department of Physics and Astronomy, University of Notre Dame, Notre Dame, IN 46556, USA}
	\affiliation{Stravropoulos Center for Complex Quantum Matter, University of Notre Dame, Notre Dame, IN 46556, USA}
    \affiliation{Center for Nanophase Materials Sciences, Oak Ridge National Laboratory, Oak Ridge, TN 37830, USA}
 	\author{Masashi Arita}
	\affiliation{Research Institute for Synchrotron Radiation Science (HISOR), Hiroshima University, Higashi-Hiroshima 739-0046, Japan}
    \author{Shin-ichiro Ideta}
	\affiliation{Research Institute for Synchrotron Radiation Science (HISOR), Hiroshima University, Higashi-Hiroshima 739-0046, Japan}
 	\author{Nirmal J. Ghimire}
	\affiliation{Department of Physics and Astronomy, University of Notre Dame, Notre Dame, IN 46556, USA}
    \affiliation{Stravropoulos Center for Complex Quantum Matter, University of Notre Dame, Notre Dame, IN 46556, USA}
    \author{Kenya Shimada}
	\affiliation{Research Institute for Synchrotron Radiation Science (HISOR), Hiroshima University, Higashi-Hiroshima 739-0046, Japan}
	\affiliation{Research Institute for Semiconductor Engineering (RISE), Hiroshima University, Higashi-Hiroshima, 739-8527, Japan}
	\affiliation{International Institute for Sustainability with Knotted Chiral Meta Matter (SKCM$^2$), Hiroshima University, Higashi-Hiroshima, 739-8531, Japan}
	\author{Anup Pradhan Sakhya}
    \thanks{Corresponding author:\href{mailto:anuppradhansakhya@hiroshima-u.ac.jp}{anuppradhansakhya@hiroshima-u.ac.jp}}\affiliation{Research Institute for Synchrotron Radiation Science (HISOR), Hiroshima University, Higashi-Hiroshima 739-0046, Japan}

\begin{abstract}

Magnetic topological semimetals provide a promising platform for emergent quantum phenomena driven by the interplay between magnetism and relativistic fermions, including anomalous transport effects and tunable topological phases. Here, we investigate the electronic structure and orbital character of EuAgAs, a magnetic topological Dirac semimetal candidate, using density functional theory (DFT) and polarization dependent angle resolved photoemission spectroscopy (ARPES). Fermi surface mapping and constant energy contours measured at 9 eV reveal ring like features that systematically expand with increasing binding energy, consistent with nearly linear low energy Dirac like dispersion. ARPES measurements at different photon energies hint at the presence of a van Hove singularity predicted by DFT calculations. Furthermore, this indicates that the photoemission matrix elements are highly sensitive to the excitation energy, allowing different photon energies to selectively probe distinct orbital characters. Polarization dependent ARPES measurements performed in $s$- and $p$-polarized geometries exhibit pronounced variations in spectral intensity, indicating symmetry selective orbital contributions to electronic states. These matrix element driven intensity modulations are well reproduced by DFT calculations. Furthermore, the observed Dirac like states remain nearly unchanged over the temperature range from 9 K to 30 K, suggesting that the magnetic ordering has minimal influence on the electronic structure. Our combined experimental and theoretical results provide detailed insight into the orbital selective electronic structure of EuAgAs and its implications for magnetic topological quantum states.

\end{abstract}

     \maketitle
     
\section{Introduction}
\begin{figure*}
     \centering
     \includegraphics[width=0.95\textwidth]{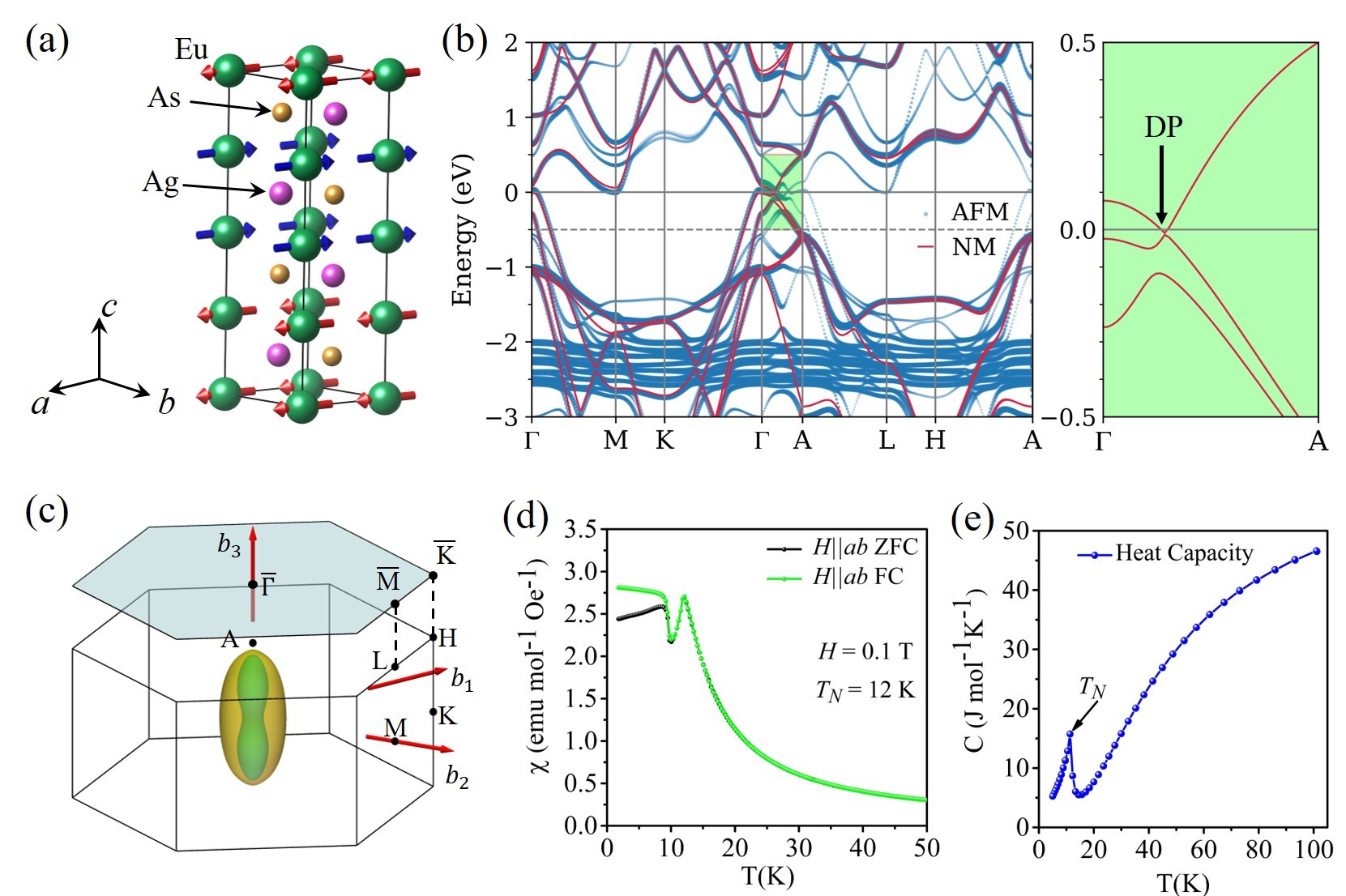}
     \caption{(a) Magnetic ground state structure of EuAgAs~\cite{gazzah2026}. (b) Calculated electronic band structures of EuAgAs, including SOC for the AFM and NM phases, are shown in blue and red, respectively. The dashed line represents the shifted Fermi level. The right panel presents a zoomed in picture of (b) along the $\Gamma$- A direction, showing the DP near the Fermi level in the NM phase. (c) Schematic representation of the bulk and surface BZs with the corresponding high symmetry points indicated. (d) Temperature dependent magnetic susceptibility of EuAgAs measured under ZFC and FC conditions with a magnetic field ($H = 0.1$ T) applied parallel to the $ab$ plane. (e) Temperature dependent heat capacity of EuAgAs showing a pronounced anomaly near $T_N = 12$ K.}
     \label{fig:1}
\end{figure*}
The experimental realization of magnetic topological quantum materials has ignited tremendous interest because of the intimate coupling between magnetism and nontrivial topological electronic states, a synergy that can dramatically reconstruct the electronic band structure and generate unprecedented quantum phenomena beyond the scope of conventional condensed matter systems \cite{bernevig2022progress, Kuroda2020, PhysRevMaterials.5.054201, PhysRevB.106.085132, zou2019study, PhysRevB.106.235119, Schrunk2022, PhysRevB.106.115112, EuSnP, gooth2019axionic, TI3, TI4, MS813, PhysRevMaterials.9.044202}. Eu-based compounds have emerged as a compelling platform for investigating the intricate connection between magnetism and topology, where strong spin orbit coupling (SOC) together with magnetism arising from highly localized 4\textit{f} electrons gives rise to a diverse range of unconventional and exotic quantum phases \cite{PhysRevB.97.214422, Du2022, Wang, PhysRevB.108.L241115, Roy, PhysRevLett.126.076602, PhysRevB.101.205126, PhysRevB.102.165153, ram2024magnetotransport, sears2025euausb, zhu2020magnetic}. EuAgAs is one such promising candidate for investigating the coexistence of nontrivial topology and magnetism. It crystallizes in the hexagonal ZrBeSi-type structure with space group $P6_3/mmc$, providing a favorable platform for exploring intertwined topological and magnetic phenomena \cite{PhysRevB.103.L241112}. Density functional theory (DFT) calculations have shown that tuning the magnetic order in EuAgAs can realize multiple magnetic topological phases, including magnetic triply degenerate points, topological mirror semimetal states, and linear as well as double Weyl points, depending on the symmetry breaking induced by the magnetic configuration \cite{PhysRevB.104.165424}. EuAgAs has been proposed as an antiferromagnetic (AFM) Dirac semimetal (DSM) that can evolve into a Weyl semimetal under magnetic-field-induced time-reversal symmetry breaking, consistent with its large topological Hall effect and chiral-anomaly-induced positive longitudinal magnetoconductivity below 12 K \cite{PhysRevB.103.L241112}. Using ultrafast optical excitation, EuAgAs was recently found to exhibit distinct fluence-dependent responses in the AFM and paramagnetic states, indicating a possible magnetic-order-driven transition between different topological phases \cite{PhysRevB.111.L121113}. Recently, detailed neutron diffraction has been performed, which reveals that EuAgAs hosts a commensurate AFM ground state with an in-plane spin arrangement, while DFT calculations show a near-degeneracy of competing magnetic phases, making the system highly tunable and potentially capable of hosting pressure-induced topological altermagnetism \cite{gazzah2026}.\\ 
\indent However, studies on EuAgAs to date have primarily relied on magnetotransport measurements and DFT calculations, with a comprehensive experimental investigation of its electronic structure using polarization dependent ARPES still lacking. Such an investigation is essential for resolving the fine details of the electronic states near the Fermi level and understanding the impact of magnetism on the band structure. In this work, we address this gap through detailed high resolution polarization dependent ARPES measurements in both the paramagnetic and AFM phases, systematically probing the evolution of the electronic structure along multiple high symmetry directions and across different \textit{k$_z$} planes, complemented by extensive DFT calculations.
Furthermore, we perform detailed polarization dependent ARPES measurements to resolve the orbital selective nature of the Dirac like states, supported by comprehensive DFT calculations with orbital projected band analysis. 

\begin{figure}[htbp]
     \centering
     \includegraphics[width=0.48\textwidth]{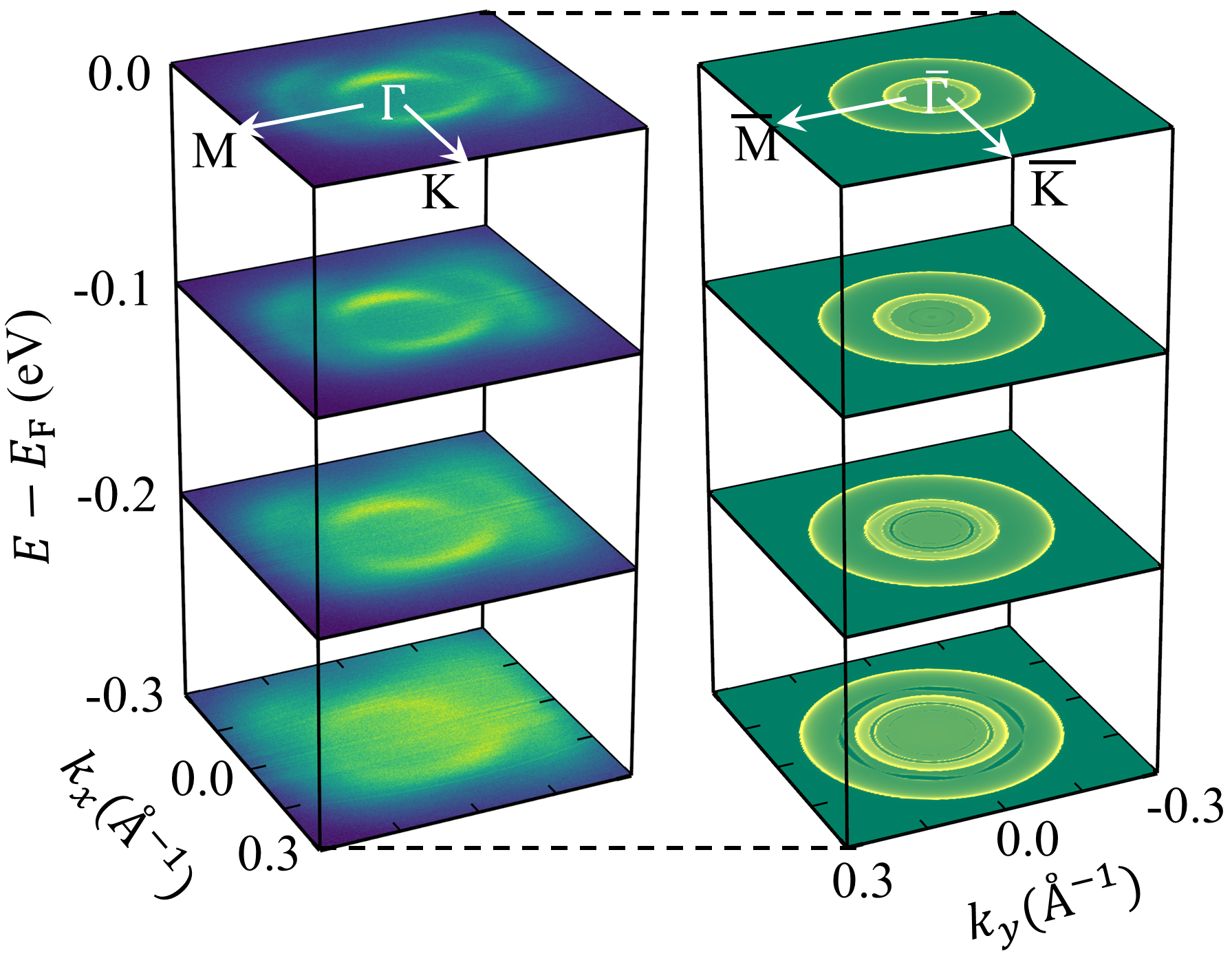}
     \caption{(Left panel) ARPES-measured Fermi surface and constant energy CECs at different binding energies acquired using a photon energy of 9 eV. The high symmetry directions are indicated. (Right panel) Corresponding DFT calculated surface projected Fermi surface and CECs showing the evolution of the electronic states with binding energy. The measurements were performed at the HiSOR BL-9A beamline at 9 K using $p$-polarized light.}
     \label{Fig:2}
\end{figure}

\section{Methods}
Magnetic properties were measured using a 7 T Quantum Design Magnetic Property Measurement System (MPMS3) operated in vibrating sample magnetometer (VSM) mode over the temperature range of 1.8–400 K, with a vibration amplitude of 8 mm and an averaging time of 2 s. Heat capacity measurements were performed using Quantum Design Dynacool Physical Property Measurement Systems (PPMS) equipped with a 9 T magnet.\\
\indent ARPES measurements were performed at the BL-9A beamline of the Research Institute for Synchrotron Radiation Science (HiSOR), Hiroshima University. The measurements were carried out using linearly polarized synchrotron radiation with photon energies ranging from 9 to 32 eV. Both $s$- and $p$-polarized geometries were employed to investigate the orbital selective electronic structure of EuAgAs. Single crystals of EuAgAs were cleaved \textit{in situ} under ultrahigh vacuum conditions better than $4 \times 10^{-9}$ Pa, exposing a clean surface for measurements. The ARPES spectra were acquired at temperatures ranging from 9 K to 30 K using a hemispherical electron analyzer (ASTRAIOS 190). The overall energy and angular resolutions were set to better than 8 meV and 0.3$^\circ$, respectively. \par
First principles calculations were carried out within the framework of DFT using the Vienna $ab$ $initio$ Simulation Package (VASP) and the projector augmented wave (PAW) method~\cite{PhysRevB.54.11169,PhysRevB.59.1758}. The exchange correlation effects were treated using the generalized gradient approximation (GGA) in the form of the Perdew–Burke–Ernzerhof functional~\cite{PhysRevLett.77.3865}, and SOC was included self consistently. A plane wave energy cutoff of 350 eV was employed throughout all nonmagnetic calculations. The Brillouin zone integration was performed using a $13 \times 13 \times 9$ $k$-mesh in the Gamma scheme. For the nonmagnetic calculations, the Eu 4$f$ electrons were treated as core states, and the Eu ion was considered in the divalent Eu$^{2+}$ configuration. To investigate the electronic topology of EuAgAs, a tight binding Hamiltonian was constructed using maximally localized Wannier functions (MLWFs) implemented in Wannier90~\cite{MOSTOFI2008685}, considering Eu-$d$, Ag-$d$, and As-$p$ orbitals as projection bases. Furthermore, the projection of bulk bands and the Fermi surface (FS) was calculated using the recursive Green’s function method as implemented in WannierTools~\cite{WU2018405}. \\
\indent The AFM electronic structure calculations were performed within the DFT+$U$ formalism with an effective on site Coulomb interaction parameter of $U=8$ eV applied to the Eu 4$f$ states. A $1 \times 1 \times 2$ supercell was employed to accommodate the magnetic ordering. Since the Eu 4$f$ electrons were treated explicitly as valence states in the AFM calculations, the plane wave energy cutoff was increased to 500 eV. In this case, the Brillouin zone (BZ) was sampled using a Gamma centered $13 \times 13 \times 5$ $k$ point mesh to account for the enlarged lattice periodicity along the $c$ axis. The calculated electronic band structure was subsequently unfolded onto the primitive BZ to enable direct comparison with the nonmagnetic calculations.

\begin{figure*}
     \centering
     \includegraphics[width=0.98\textwidth]{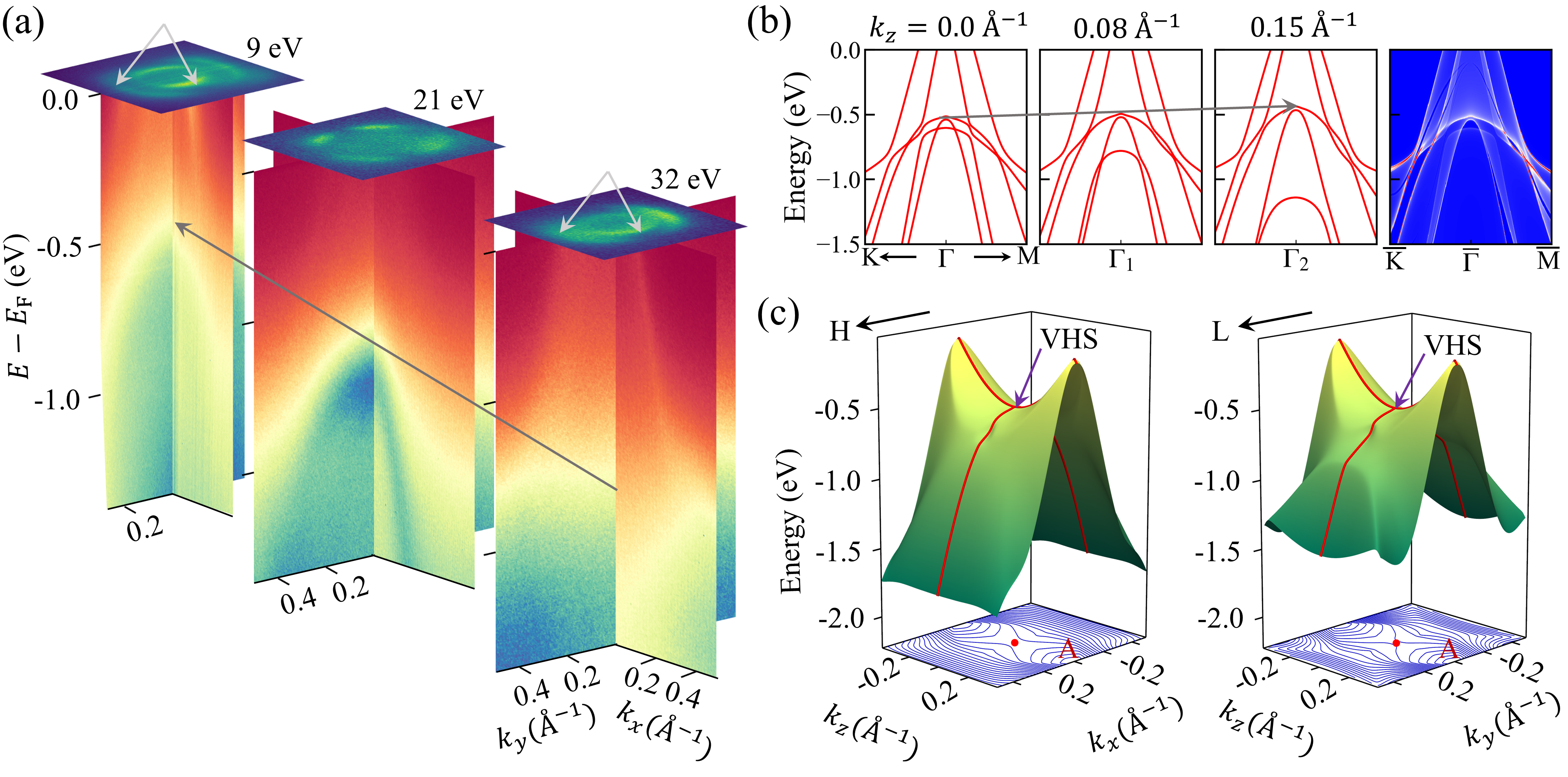}
     \caption{(a) ARPES-measured three dimensional band dispersions along the high symmetry directions M~-~$\Gamma$~-~M and K~-~$\Gamma$~-~K acquired at photon energies of 9 eV, 21 eV, and 32 eV. The corresponding Fermi surfaces measured at each photon energy are shown at the top of the three dimensional band dispersions. (b) Theoretically calculated electronic band dispersions along the K - $\Gamma$ - M direction at different $k_z$ values. The right most panel represents the surface projected band structure. (c) Theoretically calculated band dispersions in $k_x$-$k_z$ and $k_y$-$k_z$ planes, revealing the vHS at $\Gamma$.}
     \label{fig:3}
\end{figure*}

\section{Results and Discussion}

\subsection{CRYSTAL STRUCTURE, PHYSICAL PROPERTIES, AND DFT CALCULATIONS}
Figure~\ref{fig:1}(a) illustrates the crystal and magnetic structure of EuAgAs, which crystallizes in the nonsymmorphic hexagonal $P6_3/mmc$ (194) structure with alternating Eu and Ag-As layers stacked along the $c$-axis. The Eu sublattice hosts an $\uparrow\uparrow\downarrow\downarrow$ AFM ground state configuration along the $c$-direction with the magnetic moments aligned parallel to the $a$-direction, as established by neutron diffraction measurements reported by Gazzah \textit{et al}~\cite{gazzah2026}.\\
\indent To gain insight into the electronic structure of this material, we have calculated the electronic band dispersion with SOC in the AFM (structure shown in Fig.~\ref{fig:1}(a)) and nonmagnetic (NM) phases as shown in Fig.~\ref{fig:1}(b). Interestingly, close to the Fermi level, the conduction and valence bands cross along the $\Gamma$-A high symmetry direction as shown in the right panel of Fig.~\ref{fig:1}(b), forming a fourfold degenerate Dirac point (DP) protected by the $C_3$ rotational symmetry \cite{Yang2015}. The calculated band structure in the AFM phase is unfolded in the primitive nonmagnetic BZ. The linewidth and intensity of the unfolded bands are proportional to the spectral weight associated with the Bloch character of the corresponding eigenstates in the primitive cell. Consequently, bands with large spectral weight appear thick, representing the dominant itinerant electronic states, whereas weak and diffused bands originate from zone folding effects induced by the enlarged magnetic unit cell. Despite the magnetic reconstruction, the low energy Dirac like dispersive states near the Fermi level remain largely preserved in the AFM phase, indicating robust electronic topology against magnetic ordering. In addition, nearly dispersionless Eu 4$f$ states appear around -2~eV in the AFM phase, reflecting their strongly localized nature. \\
\indent Figure~\ref{fig:1}(c) presents the bulk and surface BZs of EuAgAs together with the calculated three dimensional FS and the high symmetry momentum points used in the electronic structure calculations and polarization dependent ARPES measurements. To achieve better agreement with the experimentally observed band positions and FS topology, the calculated chemical potential was shifted downward by approximately 0.5 eV (indicated by dashed line in Fig.~\ref{fig:1}(b)). Figures~\ref{fig:1}(d) and \ref{fig:1}(e) show the temperature dependent magnetic susceptibility and heat capacity measurements of EuAgAs, respectively. The magnetic susceptibility measured under both zero field cooled (ZFC) and field cooled (FC) conditions with a magnetic field applied parallel to the $ab$ plane exhibits some anomaly near 12 K, consistent with the previous reports~\cite{PhysRevB.103.L241112,gazzah2026}.

\begin{figure*}
     \centering
     \includegraphics[width=0.95\textwidth]{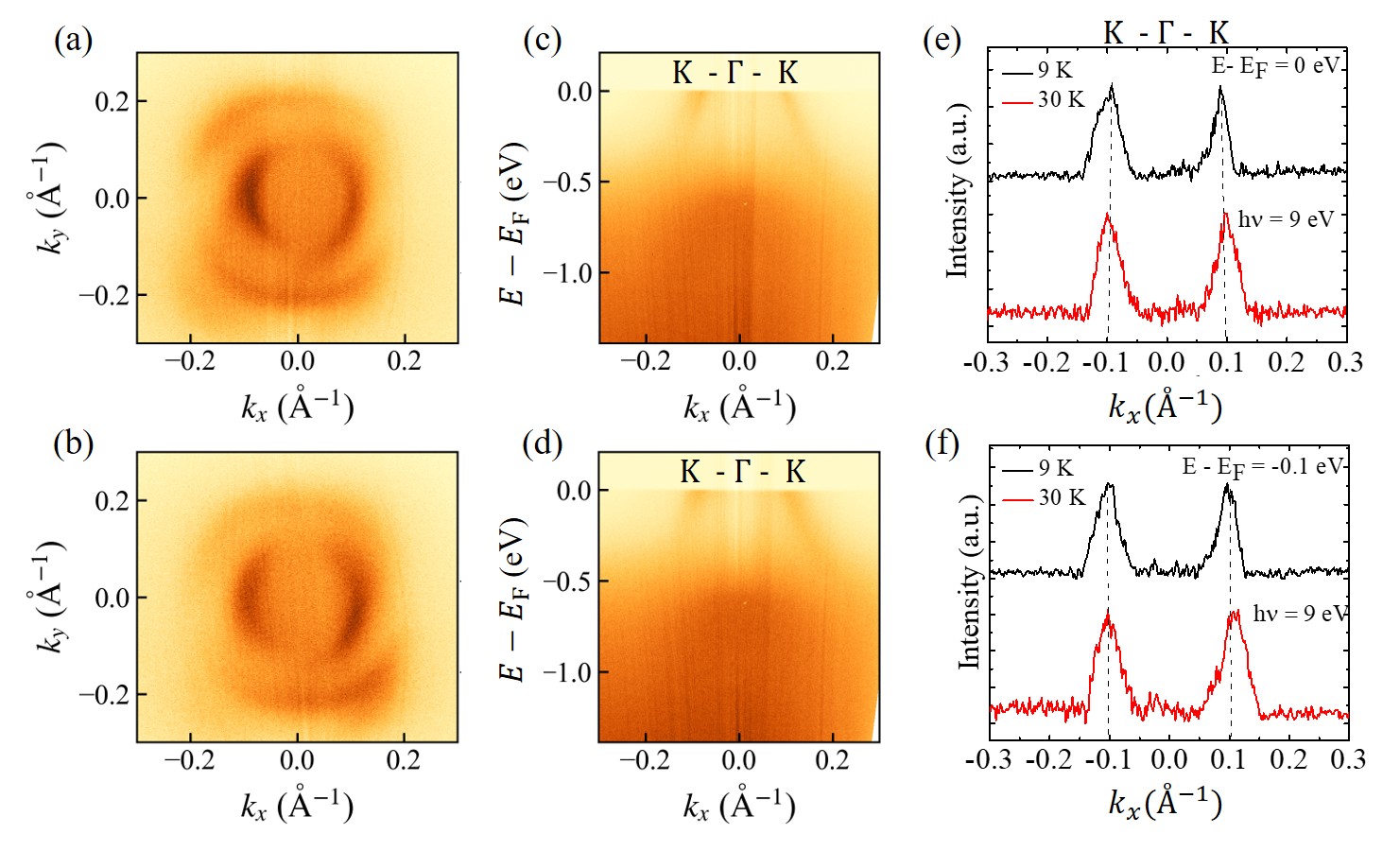}
     \caption{Temperature dependent CECs and electronic band dispersion of EuAgAs along the high symmetry direction K - $\Gamma$ - K. ARPES measured CECs at $E-E_F = 0$ eV acquired using a photon energy of 9 eV at temperatures of (a) 9 K and (b) 30 K. Experimentally measured band dispersions along the K - $\Gamma$ - K direction at (c) 9 K and (d) 30 K. MDCs along the K - $\Gamma$ - K direction at (e) $E-E_F = 0$ eV and (f) $E-E_F = -0.1$ eV, comparing the temperature evolution of the electronic structure across the AFM transition. All measurements were performed at the HiSOR BL-9A beamline using $p$-polarized light.}
     \label{fig:4}
\end{figure*} 

\subsection{FERMI SURFACE AND CONSTANT ENERGY CONTOURS}
Figure.~\ref{Fig:2} presents the experimentally measured and theoretically calculated FS and constant energy contours (CECs) of EuAgAs, providing direct insight into the low energy Dirac like electronic states near the Fermi level. The left panel shows the ARPES measured FS and CECs acquired at a photon energy of 9~eV and a temperature of 9~K using $p$-polarized light, while the right panel displays the corresponding surface projected CECs obtained from DFT calculations at identical binding energies.
The ARPES measurements reveal a well defined dispersive band near the Fermi level, forming ring like contours centered around the $\Gamma$ point that systematically expand with increasing binding energy. This evolution indicates the presence of nearly linearly dispersive electronic states near $E_F$. Notably, the ARPES intensity exhibits a pronounced anisotropic spectral weight distribution along the $\Gamma$-M and $\Gamma$-K directions (Fig.~\ref{Fig:2}), indicating strong matrix element effects and orbital selective contributions to the low energy states \cite{moser2017experimentalist, zhang2022angle}. The observed anisotropic intensity distribution, despite the underlying hexagonal crystal symmetry, indicates that the low energy spectral weight is strongly governed by photoemission matrix element effects rather than intrinsic symmetry breaking of the electronic structure. Such intensity modulation suggests the presence of distinct orbital characters participating in the Dirac like bands, which will be discussed in detail in the polarization dependent ARPES section.

\subsection{PHOTON ENERGY DEPENDENT BAND STRUCTURE}

To understand the evolution of the electronic structure with $k_z$, \textit{i.e.}, the three dimensional nature of the bands in EuAgAs, we performed photon energy dependent ARPES measurements. Fig. \ref{fig:3}(a) presents the experimentally measured band dispersions along the high symmetry directions M $-\Gamma–$ M and K $-\Gamma–$ K using $p$-polarized light at photon energies of 9 and 32 eV, and using $s$-polarized light at photon energy 21 eV, which correspond to different $k_z$ planes in momentum space. The corresponding FS maps measured at each photon energy are displayed on top of the respective band dispersions to visualize the evolution of FS with $k_z$. Interestingly, although the overall FS shape shows minimal variation with photon energy, consistent with the relatively small $k_z$-dependence predicted by DFT calculations (see Supplementary Information~\cite{Suppl}), the spectral intensity distribution undergoes substantial redistribution. In particular, the intensity around the ring, like Fermi contours, varies strongly as a function of photon energy. Such behavior indicates that different photon energies selectively enhance distinct orbital components of the low energy electronic states via photoemission matrix element effects \cite{moser2017experimentalist}. A detailed discussion of the orbital character and polarization dependent electronic structure is presented in the section \ref{pol}. \par
Fig.~\ref{fig:3}(b) presents the theoretically calculated band dispersions along the K~-$\Gamma$-~M direction at different $k_z$ values to understand the experimentally observed $k_z$-dependent evolution. The right most panel of Fig.~\ref{fig:3}(b), showing the surface projected spectral function, qualitatively reproduces the experimentally observed bands. With increasing $k_z$, the upper portion of the hole like band gradually shifts toward higher energy, as indicated by the arrow in Fig.~\ref{fig:3}(b). This evolution signifies that the curvature of the band along the $k_z$ direction is opposite to that along the in plane $k_x$ and $k_y$ directions. Consequently, the band develops a saddle point like topology at the $\Gamma$ point, giving rise to a van Hove Singularity (vHS). The ARPES spectra also exhibit a similar upward evolution of the hole like dispersion with changing photon energy, consistent with the theoretically predicted vHS related band evolution \cite{hossain2025superconductivity}. 
To further visualize the vHS, Fig.~\ref{fig:3}(c) presents the calculated three dimensional band dispersions in the $k_x$-$k_z$ and $k_y$-$k_z$ momentum planes. The red dots mark the locations of the vHS point, and the color gradient represents energy. Notably, the observed vHS belongs to the higher order vHS class \cite{hu2022rich}, characterized by an extended flat dispersion near the saddle point, which can significantly enhance electronic instabilities and correlation effects (see Supplementary Information~\cite{Suppl}).

\begin{figure*}
     \centering
     \includegraphics[width=0.98\textwidth]{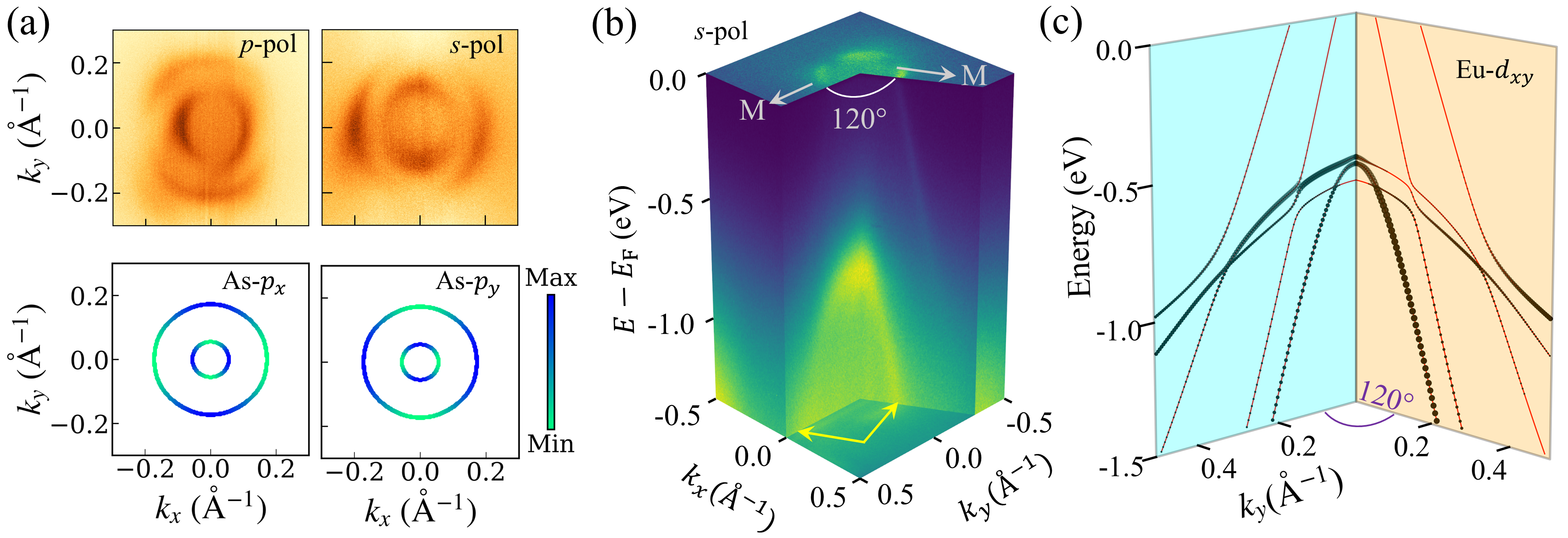}
     \caption{(a) FS maps measured using $p$-polarization and $s$-polarization geometries, where the upper panels show experimental results. The DFT calculated orbital contribution from As-$p$ is shown in the lower panel. (b) ARPES intensity map measured using 21 eV photons with $s$-polarized light along the two $\Gamma$–M directions, showing a pronounced intensity variation between the equivalent momentum cuts. (c) Theoretically calculated band dispersion along M–$\Gamma$–M with the projected Eu-$d_{xy}$ orbital contribution, explaining the anisotropic spectral intensity observed in the ARPES measurements.}
     \label{fig:5}
\end{figure*}

\subsection{TEMPERATURE DEPENDENT BAND STRUCTURE}

To investigate the influence of AFM ordering on the electronic structure of EuAgAs, we have performed temperature dependent ARPES measurements using a photon energy of 9 eV, as shown in Fig.~\ref{fig:4}. The measurements were carried out at 9 K and 30 K, corresponding to temperatures below and above the Néel temperature, respectively. The FS maps measured below and above the Néel temperature are presented in Figs.~\ref{fig:4}(a) and \ref{fig:4}(b), while the corresponding ARPES band dispersions along the K~-$\Gamma$-~K direction are shown in Figs.~\ref{fig:4}(c) and \ref{fig:4}(d). A comparison of the FS shape and band dispersions does not reveal any observable changes in the low energy electronic structure across the magnetic transition with temperature.
To quantitatively examine possible temperature dependent changes, we analyzed the momentum distribution curves (MDCs) along the K~-$\Gamma$-~K direction at $E-E_F = 0$ eV and $E-E_F = -0.1$ eV, which are shown in Figs.~\ref{fig:4}(e) and \ref{fig:4}(f), respectively. The MDC profiles exhibit nearly identical peak positions and spectral features at 9 K and 30 K, without any detectable momentum shift or spectral reconstruction. \\
\indent The absence of observable changes in the ARPES spectra across the AFM transition suggests that magnetic ordering has only a weak influence on the low energy electronic structure of EuAgAs. This behavior can be attributed to the localized nature of the Eu 4$f$ moments, whose electronic states lie at higher binding energies, leading to weak coupling with the low energy states near the Fermi level \cite{EuSnP}. In contrast, the low energy electronic structure is predominantly derived from highly dispersive As and Ag $sp$ states, for which the magnetic reconstruction is expected to be relatively weak. The detailed polarization dependent ARPES measurements have been discussed in the next section. Furthermore, the calculated electronic band structures for the AFM and NM phases shown in Fig.~\ref{fig:1}(b) exhibit only minimal differences in the vicinity of $E_F$, indicating the absence of significant magnetic reconstruction of the low energy band topology. 


\subsection{POLARIZATION DEPENDENT ELECTRONIC STRUCTURE}
\label{pol}
To elucidate the orbital character of the low energy electronic states, we performed polarization dependent ARPES measurements. In the present experimental geometry, the sample surface lies in the $xy$ plane, while the incident photons propagate along the $x$ direction. The $xz$ plane, defined by the incident beam and the analyzer slit, therefore acts as the mirror plane. According to the dipole selection rules in ARPES \cite{PhysRevB.100.205135, moser2017experimentalist}, $p$ polarized light predominantly probes even orbitals with respect to the mirror plane, while $s$ polarized light selectively enhances odd orbitals.\\
\indent Fig.~\ref{fig:5}(a) presents the FS measured using 9 eV photon energy under both $s$ and $p$ polarized light configurations, revealing a nearly circular pocket centered at the $\Gamma$ point. Although the overall FS shape remains nearly unchanged, the spectral intensity distribution exhibits a pronounced variation between the $p$ and $s$ polarized geometries. Under $p$ polarized light, the spectral weight for the outer pocket is strongly enhanced along the vertical FS segments, whereas under $s$ polarized light, the intensity becomes dominant along the horizontal portions of the FS contour. Such complementary intensity modulation is a hallmark of orbital selective photoemission arising from dipole selection rules \cite{moser2017experimentalist}. 
The lower panels of Fig.~\ref{fig:5}(a) present the orbital projected DFT FS calculations for As-$p_x$ and As-$p_y$ orbitals. The calculated As-$p_x$ projection reproduces the intensity pattern observed experimentally in the $p$ polarized geometry, confirming that the dominant contribution originates from the even symmetry As-$p_x$ orbital. In contrast, the As-$p_y$ orbital projection closely matches the spectral intensity distribution observed under $s$ polarized light, indicating its odd symmetry character.\\
\indent Fig.~\ref{fig:5}(b) shows the ARPES intensity map measured using 21 eV photons under $s$ polarized light. Compared with the 9 eV measurements, the overall FS shape remains nearly unchanged, but the spectral intensity distribution changes dramatically. In particular, the spectral weight becomes predominantly concentrated at the four corners of the FS. Such a pronounced photon energy dependent intensity variation indicates that varying the photon energy selectively enhances different orbital contributions to the same FS. To identify the orbital origin of these states, the experimentally observed band dispersions are compared with orbital resolved DFT calculations (see Supplementary Information~\cite{Suppl}). The comparison reveals that the FS states strongly enhanced at 21 eV predominantly originate from the Eu-$d_{yz}$ orbital. To further demonstrate the orbital selectivity, the M-$\Gamma$-M band dispersion is presented in Fig.~\ref{fig:5}(c), where different bands at higher binding energy are selectively enhanced along the same momentum direction. These bands are mainly derived from the Eu-$d_{xy}$ orbital, which possesses odd symmetry with respect to the experimental mirror plane, as evidenced by the close correspondence between the experimental dispersions and the calculated orbital projections (Fig.~\ref{fig:5}(c)).

\section{Conclusion}

In conclusion, we have investigated the electronic structure of the AFM topological semimetal candidate EuAgAs using polarization, photon energy, and temperature dependent ARPES measurements, combined with first principles calculations. The ARPES measurements reveal Dirac like dispersive electronic states near the Fermi level, whose evolution is qualitatively consistent with the calculated electronic structure. Photon energy dependent measurements demonstrate the three dimensional nature of the low energy bands and reveal the presence of a higher order vHS arising from a saddle point like band topology near the $\Gamma$ point. Polarization dependent ARPES measurements further reveal pronounced orbital selective spectral weight modulation driven by photoemission matrix element effects, supported by orbital resolved DFT calculations. Despite the bulk AFM ordering below $T_N = 12$ K, the low energy electronic structure remains essentially unchanged across the magnetic transition, indicating weak coupling between the localized Eu 4$f$ moments and the electronic states near the Fermi level. Our results establish EuAgAs as a promising platform for exploring the interplay between topology, orbital selective electronic structure, and magnetism in correlated quantum materials.


\begin{acknowledgements}
The ARPES measurements were performed with the approval of the Proposal Assessing Committee of the Research Institute for Synchrotron Radiation Science (Proposal No. 25BU015). We thank N-BARD, Hiroshima University, for supplying the liquid helium. The work at Hiroshima University was supported by JST EXPERT-J, Japan Grant Number JPMJEX2510. N.J.G acknowledges the support of the NSF CAREER award DMR-2143903.
\end{acknowledgements}

\vspace{2ex}
\textit{Note added.} During the preparation of this manuscript, a related work was published on arXiv reporting helical magnetic ordering, in which the transverse spin helix induces band folding and gives rise to higher-order Weyl nodes. \cite{soh2026}.

\vspace{2ex}

\noindent\textbf{Competing interests}

\noindent The authors declare no competing interests.
\vspace{2ex}

\noindent\textbf{ADDITIONAL INFORMATION}\\
Correspondence and requests for materials should be addressed to Anup Pradhan Sakhya.

\bigskip

\begin{thebibliography}{42}%
\makeatletter
\providecommand \@ifxundefined [1]{%
 \@ifx{#1\undefined}
}%
\providecommand \@ifnum [1]{%
 \ifnum #1\expandafter \@firstoftwo
 \else \expandafter \@secondoftwo
 \fi
}%
\providecommand \@ifx [1]{%
 \ifx #1\expandafter \@firstoftwo
 \else \expandafter \@secondoftwo
 \fi
}%
\providecommand \natexlab [1]{#1}%
\providecommand \enquote  [1]{``#1''}%
\providecommand \bibnamefont  [1]{#1}%
\providecommand \bibfnamefont [1]{#1}%
\providecommand \citenamefont [1]{#1}%
\providecommand \href@noop [0]{\@secondoftwo}%
\providecommand \href [0]{\begingroup \@sanitize@url \@href}%
\providecommand \@href[1]{\@@startlink{#1}\@@href}%
\providecommand \@@href[1]{\endgroup#1\@@endlink}%
\providecommand \@sanitize@url [0]{\catcode `\\12\catcode `\$12\catcode `\&12\catcode `\#12\catcode `\^12\catcode `\_12\catcode `\%12\relax}%
\providecommand \@@startlink[1]{}%
\providecommand \@@endlink[0]{}%
\providecommand \url  [0]{\begingroup\@sanitize@url \@url }%
\providecommand \@url [1]{\endgroup\@href {#1}{\urlprefix }}%
\providecommand \urlprefix  [0]{URL }%
\providecommand \Eprint [0]{\href }%
\providecommand \doibase [0]{https://doi.org/}%
\providecommand \selectlanguage [0]{\@gobble}%
\providecommand \bibinfo  [0]{\@secondoftwo}%
\providecommand \bibfield  [0]{\@secondoftwo}%
\providecommand \translation [1]{[#1]}%
\providecommand \BibitemOpen [0]{}%
\providecommand \bibitemStop [0]{}%
\providecommand \bibitemNoStop [0]{.\EOS\space}%
\providecommand \EOS [0]{\spacefactor3000\relax}%
\providecommand \BibitemShut  [1]{\csname bibitem#1\endcsname}%
\let\auto@bib@innerbib\@empty
\bibitem [{\citenamefont {Bernevig}\ \emph {et~al.}(2022)\citenamefont {Bernevig}, \citenamefont {Felser},\ and\ \citenamefont {Beidenkopf}}]{bernevig2022progress}%
  \BibitemOpen
  \bibfield  {author} {\bibinfo {author} {\bibfnamefont {B.~A.}\ \bibnamefont {Bernevig}}, \bibinfo {author} {\bibfnamefont {C.}~\bibnamefont {Felser}},\ and\ \bibinfo {author} {\bibfnamefont {H.}~\bibnamefont {Beidenkopf}},\ }\bibfield  {title} {\bibinfo {title} {Progress and prospects in magnetic topological materials},\ }\href {https://doi.org/https://doi.org/10.1038/s41586-021-04105-x} {\bibfield  {journal} {\bibinfo  {journal} {Nature}\ }\textbf {\bibinfo {volume} {603}},\ \bibinfo {pages} {41} (\bibinfo {year} {2022})}\BibitemShut {NoStop}%
\bibitem [{\citenamefont {Kuroda}\ \emph {et~al.}(2020)\citenamefont {Kuroda}, \citenamefont {Arai}, \citenamefont {Rezaei}, \citenamefont {Kunisada}, \citenamefont {Sakuragi}, \citenamefont {Alaei}, \citenamefont {Kinoshita}, \citenamefont {Bareille}, \citenamefont {Noguchi}, \citenamefont {Nakayama}, \citenamefont {Akebi}, \citenamefont {Sakano}, \citenamefont {Kawaguchi}, \citenamefont {Arita}, \citenamefont {Ideta}, \citenamefont {Tanaka}, \citenamefont {Kitazawa}, \citenamefont {Okazaki}, \citenamefont {Tokunaga}, \citenamefont {Haga}, \citenamefont {Shin}, \citenamefont {Suzuki}, \citenamefont {Arita},\ and\ \citenamefont {Kondo}}]{Kuroda2020}%
  \BibitemOpen
  \bibfield  {author} {\bibinfo {author} {\bibfnamefont {K.}~\bibnamefont {Kuroda}}, \bibinfo {author} {\bibfnamefont {Y.}~\bibnamefont {Arai}}, \bibinfo {author} {\bibfnamefont {N.}~\bibnamefont {Rezaei}}, \bibinfo {author} {\bibfnamefont {S.}~\bibnamefont {Kunisada}}, \bibinfo {author} {\bibfnamefont {S.}~\bibnamefont {Sakuragi}}, \bibinfo {author} {\bibfnamefont {M.}~\bibnamefont {Alaei}}, \bibinfo {author} {\bibfnamefont {Y.}~\bibnamefont {Kinoshita}}, \bibinfo {author} {\bibfnamefont {C.}~\bibnamefont {Bareille}}, \bibinfo {author} {\bibfnamefont {R.}~\bibnamefont {Noguchi}}, \bibinfo {author} {\bibfnamefont {M.}~\bibnamefont {Nakayama}}, \bibinfo {author} {\bibfnamefont {S.}~\bibnamefont {Akebi}}, \bibinfo {author} {\bibfnamefont {M.}~\bibnamefont {Sakano}}, \bibinfo {author} {\bibfnamefont {K.}~\bibnamefont {Kawaguchi}}, \bibinfo {author} {\bibfnamefont {M.}~\bibnamefont {Arita}}, \bibinfo {author} {\bibfnamefont {S.}~\bibnamefont {Ideta}}, \bibinfo {author} {\bibfnamefont {K.}~\bibnamefont {Tanaka}},
  \bibinfo {author} {\bibfnamefont {H.}~\bibnamefont {Kitazawa}}, \bibinfo {author} {\bibfnamefont {K.}~\bibnamefont {Okazaki}}, \bibinfo {author} {\bibfnamefont {M.}~\bibnamefont {Tokunaga}}, \bibinfo {author} {\bibfnamefont {Y.}~\bibnamefont {Haga}}, \bibinfo {author} {\bibfnamefont {S.}~\bibnamefont {Shin}}, \bibinfo {author} {\bibfnamefont {H.~S.}\ \bibnamefont {Suzuki}}, \bibinfo {author} {\bibfnamefont {R.}~\bibnamefont {Arita}},\ and\ \bibinfo {author} {\bibfnamefont {T.}~\bibnamefont {Kondo}},\ }\bibfield  {title} {\bibinfo {title} {Devil's staircase transition of the electronic structures in {CeSb}},\ }\href {https://doi.org/10.1038/s41467-020-16707-6} {\bibfield  {journal} {\bibinfo  {journal} {Nat. Commun.}\ }\textbf {\bibinfo {volume} {11}},\ \bibinfo {pages} {2888} (\bibinfo {year} {2020})}\BibitemShut {NoStop}%
\bibitem [{\citenamefont {Sakhya}\ \emph {et~al.}(2021)\citenamefont {Sakhya}, \citenamefont {Paulose}, \citenamefont {Thamizhavel},\ and\ \citenamefont {Maiti}}]{PhysRevMaterials.5.054201}%
  \BibitemOpen
  \bibfield  {author} {\bibinfo {author} {\bibfnamefont {A.~P.}\ \bibnamefont {Sakhya}}, \bibinfo {author} {\bibfnamefont {P.~L.}\ \bibnamefont {Paulose}}, \bibinfo {author} {\bibfnamefont {A.}~\bibnamefont {Thamizhavel}},\ and\ \bibinfo {author} {\bibfnamefont {K.}~\bibnamefont {Maiti}},\ }\bibfield  {title} {\bibinfo {title} {Evidence of nontrivial {B}erry phase and {K}ondo physics in {SmBi}},\ }\href {https://doi.org/10.1103/PhysRevMaterials.5.054201} {\bibfield  {journal} {\bibinfo  {journal} {Phys. Rev. Mater.}\ }\textbf {\bibinfo {volume} {5}},\ \bibinfo {pages} {054201} (\bibinfo {year} {2021})}\BibitemShut {NoStop}%
\bibitem [{\citenamefont {Sakhya}\ \emph {et~al.}(2022{\natexlab{a}})\citenamefont {Sakhya}, \citenamefont {Kumar}, \citenamefont {Pramanik}, \citenamefont {Pandeya}, \citenamefont {Verma}, \citenamefont {Singh}, \citenamefont {Datta}, \citenamefont {Sasmal}, \citenamefont {Mondal}, \citenamefont {Schwier}, \citenamefont {Shimada}, \citenamefont {Thamizhavel},\ and\ \citenamefont {Maiti}}]{PhysRevB.106.085132}%
  \BibitemOpen
  \bibfield  {author} {\bibinfo {author} {\bibfnamefont {A.~P.}\ \bibnamefont {Sakhya}}, \bibinfo {author} {\bibfnamefont {S.}~\bibnamefont {Kumar}}, \bibinfo {author} {\bibfnamefont {A.}~\bibnamefont {Pramanik}}, \bibinfo {author} {\bibfnamefont {R.~P.}\ \bibnamefont {Pandeya}}, \bibinfo {author} {\bibfnamefont {R.}~\bibnamefont {Verma}}, \bibinfo {author} {\bibfnamefont {B.}~\bibnamefont {Singh}}, \bibinfo {author} {\bibfnamefont {S.}~\bibnamefont {Datta}}, \bibinfo {author} {\bibfnamefont {S.}~\bibnamefont {Sasmal}}, \bibinfo {author} {\bibfnamefont {R.}~\bibnamefont {Mondal}}, \bibinfo {author} {\bibfnamefont {E.~F.}\ \bibnamefont {Schwier}}, \bibinfo {author} {\bibfnamefont {K.}~\bibnamefont {Shimada}}, \bibinfo {author} {\bibfnamefont {A.}~\bibnamefont {Thamizhavel}},\ and\ \bibinfo {author} {\bibfnamefont {K.}~\bibnamefont {Maiti}},\ }\bibfield  {title} {\bibinfo {title} {Behavior of gapped and ungapped {D}irac cones in the antiferromagnetic topological metal {SmBi}},\ }\href
  {https://doi.org/10.1103/PhysRevB.106.085132} {\bibfield  {journal} {\bibinfo  {journal} {Phys. Rev. B}\ }\textbf {\bibinfo {volume} {106}},\ \bibinfo {pages} {085132} (\bibinfo {year} {2022}{\natexlab{a}})}\BibitemShut {NoStop}%
\bibitem [{\citenamefont {Zou}\ \emph {et~al.}(2019)\citenamefont {Zou}, \citenamefont {He},\ and\ \citenamefont {Xu}}]{zou2019study}%
  \BibitemOpen
  \bibfield  {author} {\bibinfo {author} {\bibfnamefont {J.}~\bibnamefont {Zou}}, \bibinfo {author} {\bibfnamefont {Z.}~\bibnamefont {He}},\ and\ \bibinfo {author} {\bibfnamefont {G.}~\bibnamefont {Xu}},\ }\bibfield  {title} {\bibinfo {title} {The study of magnetic topological semimetals by first principles calculations},\ }\href {https://doi.org/https://doi.org/10.1038/s41524-019-0237-5} {\bibfield  {journal} {\bibinfo  {journal} {npj Comput. Mater.}\ }\textbf {\bibinfo {volume} {5}},\ \bibinfo {pages} {96} (\bibinfo {year} {2019})}\BibitemShut {NoStop}%
\bibitem [{\citenamefont {Sakhya}\ \emph {et~al.}(2022{\natexlab{b}})\citenamefont {Sakhya}, \citenamefont {Wang}, \citenamefont {Kabir}, \citenamefont {Huang}, \citenamefont {Hosen}, \citenamefont {Singh}, \citenamefont {Regmi}, \citenamefont {Dhakal}, \citenamefont {Dimitri}, \citenamefont {Sprague}, \citenamefont {Smith}, \citenamefont {Bauer}, \citenamefont {Ronning}, \citenamefont {Bansil},\ and\ \citenamefont {Neupane}}]{PhysRevB.106.235119}%
  \BibitemOpen
  \bibfield  {author} {\bibinfo {author} {\bibfnamefont {A.~P.}\ \bibnamefont {Sakhya}}, \bibinfo {author} {\bibfnamefont {B.}~\bibnamefont {Wang}}, \bibinfo {author} {\bibfnamefont {F.}~\bibnamefont {Kabir}}, \bibinfo {author} {\bibfnamefont {C.-Y.}\ \bibnamefont {Huang}}, \bibinfo {author} {\bibfnamefont {M.~M.}\ \bibnamefont {Hosen}}, \bibinfo {author} {\bibfnamefont {B.}~\bibnamefont {Singh}}, \bibinfo {author} {\bibfnamefont {S.}~\bibnamefont {Regmi}}, \bibinfo {author} {\bibfnamefont {G.}~\bibnamefont {Dhakal}}, \bibinfo {author} {\bibfnamefont {K.}~\bibnamefont {Dimitri}}, \bibinfo {author} {\bibfnamefont {M.}~\bibnamefont {Sprague}}, \bibinfo {author} {\bibfnamefont {R.}~\bibnamefont {Smith}}, \bibinfo {author} {\bibfnamefont {E.~D.}\ \bibnamefont {Bauer}}, \bibinfo {author} {\bibfnamefont {F.}~\bibnamefont {Ronning}}, \bibinfo {author} {\bibfnamefont {A.}~\bibnamefont {Bansil}},\ and\ \bibinfo {author} {\bibfnamefont {M.}~\bibnamefont {Neupane}},\ }\bibfield  {title} {\bibinfo {title} {Complex
  electronic structure evolution of {NdSb} across the magnetic transition},\ }\href {https://doi.org/10.1103/PhysRevB.106.235119} {\bibfield  {journal} {\bibinfo  {journal} {Phys. Rev. B}\ }\textbf {\bibinfo {volume} {106}},\ \bibinfo {pages} {235119} (\bibinfo {year} {2022}{\natexlab{b}})}\BibitemShut {NoStop}%
\bibitem [{\citenamefont {Schrunk}\ \emph {et~al.}(2022)\citenamefont {Schrunk}, \citenamefont {Kushnirenko}, \citenamefont {Kuthanazhi}, \citenamefont {Ahn}, \citenamefont {Wang}, \citenamefont {O'Leary}, \citenamefont {Lee}, \citenamefont {Eaton}, \citenamefont {Fedorov}, \citenamefont {Lou}, \citenamefont {Voroshnin}, \citenamefont {Clark}, \citenamefont {S{\'a}nchez-Barriga}, \citenamefont {Bud'ko}, \citenamefont {Slager}, \citenamefont {Canfield},\ and\ \citenamefont {Kaminski}}]{Schrunk2022}%
  \BibitemOpen
  \bibfield  {author} {\bibinfo {author} {\bibfnamefont {B.}~\bibnamefont {Schrunk}}, \bibinfo {author} {\bibfnamefont {Y.}~\bibnamefont {Kushnirenko}}, \bibinfo {author} {\bibfnamefont {B.}~\bibnamefont {Kuthanazhi}}, \bibinfo {author} {\bibfnamefont {J.}~\bibnamefont {Ahn}}, \bibinfo {author} {\bibfnamefont {L.-L.}\ \bibnamefont {Wang}}, \bibinfo {author} {\bibfnamefont {E.}~\bibnamefont {O'Leary}}, \bibinfo {author} {\bibfnamefont {K.}~\bibnamefont {Lee}}, \bibinfo {author} {\bibfnamefont {A.}~\bibnamefont {Eaton}}, \bibinfo {author} {\bibfnamefont {A.}~\bibnamefont {Fedorov}}, \bibinfo {author} {\bibfnamefont {R.}~\bibnamefont {Lou}}, \bibinfo {author} {\bibfnamefont {V.}~\bibnamefont {Voroshnin}}, \bibinfo {author} {\bibfnamefont {O.~J.}\ \bibnamefont {Clark}}, \bibinfo {author} {\bibfnamefont {J.}~\bibnamefont {S{\'a}nchez-Barriga}}, \bibinfo {author} {\bibfnamefont {S.~L.}\ \bibnamefont {Bud'ko}}, \bibinfo {author} {\bibfnamefont {R.-J.}\ \bibnamefont {Slager}}, \bibinfo {author} {\bibfnamefont
  {P.~C.}\ \bibnamefont {Canfield}},\ and\ \bibinfo {author} {\bibfnamefont {A.}~\bibnamefont {Kaminski}},\ }\bibfield  {title} {\bibinfo {title} {Emergence of {F}ermi arcs due to magnetic splitting in an antiferromagnet},\ }\href {https://doi.org/10.1038/s41586-022-04412-x} {\bibfield  {journal} {\bibinfo  {journal} {Nature}\ }\textbf {\bibinfo {volume} {603}},\ \bibinfo {pages} {610} (\bibinfo {year} {2022})}\BibitemShut {NoStop}%
\bibitem [{\citenamefont {Kushnirenko}\ \emph {et~al.}(2022)\citenamefont {Kushnirenko}, \citenamefont {Schrunk}, \citenamefont {Kuthanazhi}, \citenamefont {Wang}, \citenamefont {Ahn}, \citenamefont {O'Leary}, \citenamefont {Eaton}, \citenamefont {Bud'ko}, \citenamefont {Slager}, \citenamefont {Canfield},\ and\ \citenamefont {Kaminski}}]{PhysRevB.106.115112}%
  \BibitemOpen
  \bibfield  {author} {\bibinfo {author} {\bibfnamefont {Y.}~\bibnamefont {Kushnirenko}}, \bibinfo {author} {\bibfnamefont {B.}~\bibnamefont {Schrunk}}, \bibinfo {author} {\bibfnamefont {B.}~\bibnamefont {Kuthanazhi}}, \bibinfo {author} {\bibfnamefont {L.-L.}\ \bibnamefont {Wang}}, \bibinfo {author} {\bibfnamefont {J.}~\bibnamefont {Ahn}}, \bibinfo {author} {\bibfnamefont {E.}~\bibnamefont {O'Leary}}, \bibinfo {author} {\bibfnamefont {A.}~\bibnamefont {Eaton}}, \bibinfo {author} {\bibfnamefont {S.~L.}\ \bibnamefont {Bud'ko}}, \bibinfo {author} {\bibfnamefont {R.-J.}\ \bibnamefont {Slager}}, \bibinfo {author} {\bibfnamefont {P.~C.}\ \bibnamefont {Canfield}},\ and\ \bibinfo {author} {\bibfnamefont {A.}~\bibnamefont {Kaminski}},\ }\bibfield  {title} {\bibinfo {title} {Rare-earth monopnictides: {F}amily of antiferromagnets hosting magnetic {F}ermi arcs},\ }\href {https://doi.org/10.1103/PhysRevB.106.115112} {\bibfield  {journal} {\bibinfo  {journal} {Phys. Rev. B}\ }\textbf {\bibinfo {volume} {106}},\ \bibinfo
  {pages} {115112} (\bibinfo {year} {2022})}\BibitemShut {NoStop}%
\bibitem [{\citenamefont {Sprague}\ \emph {et~al.}(2024)\citenamefont {Sprague}, \citenamefont {Sakhya}, \citenamefont {Regmi}, \citenamefont {Mondal}, \citenamefont {Elius}, \citenamefont {Valadez}, \citenamefont {Booth}, \citenamefont {Romanova}, \citenamefont {Ptok}, \citenamefont {Kaczorowski} \emph {et~al.}}]{EuSnP}%
  \BibitemOpen
  \bibfield  {author} {\bibinfo {author} {\bibfnamefont {M.}~\bibnamefont {Sprague}}, \bibinfo {author} {\bibfnamefont {A.~P.}\ \bibnamefont {Sakhya}}, \bibinfo {author} {\bibfnamefont {S.}~\bibnamefont {Regmi}}, \bibinfo {author} {\bibfnamefont {M.~I.}\ \bibnamefont {Mondal}}, \bibinfo {author} {\bibfnamefont {I.~B.}\ \bibnamefont {Elius}}, \bibinfo {author} {\bibfnamefont {N.}~\bibnamefont {Valadez}}, \bibinfo {author} {\bibfnamefont {K.}~\bibnamefont {Booth}}, \bibinfo {author} {\bibfnamefont {T.}~\bibnamefont {Romanova}}, \bibinfo {author} {\bibfnamefont {A.}~\bibnamefont {Ptok}}, \bibinfo {author} {\bibfnamefont {D.}~\bibnamefont {Kaczorowski}}, \emph {et~al.},\ }\bibfield  {title} {\bibinfo {title} {Complex fermiology and electronic structure of antiferromagnet {EuSnP}},\ }\href {https://doi.org/https://doi.org/10.1103/PhysRevMaterials.8.054411} {\bibfield  {journal} {\bibinfo  {journal} {Phys. Rev. Mater.}\ }\textbf {\bibinfo {volume} {8}},\ \bibinfo {pages} {054411} (\bibinfo {year}
  {2024})}\BibitemShut {NoStop}%
\bibitem [{\citenamefont {Gooth}\ \emph {et~al.}(2019)\citenamefont {Gooth}, \citenamefont {Bradlyn}, \citenamefont {Honnali}, \citenamefont {Schindler}, \citenamefont {Kumar}, \citenamefont {Noky}, \citenamefont {Qi}, \citenamefont {Shekhar}, \citenamefont {Sun}, \citenamefont {Wang} \emph {et~al.}}]{gooth2019axionic}%
  \BibitemOpen
  \bibfield  {author} {\bibinfo {author} {\bibfnamefont {J.}~\bibnamefont {Gooth}}, \bibinfo {author} {\bibfnamefont {B.}~\bibnamefont {Bradlyn}}, \bibinfo {author} {\bibfnamefont {S.}~\bibnamefont {Honnali}}, \bibinfo {author} {\bibfnamefont {C.}~\bibnamefont {Schindler}}, \bibinfo {author} {\bibfnamefont {N.}~\bibnamefont {Kumar}}, \bibinfo {author} {\bibfnamefont {J.}~\bibnamefont {Noky}}, \bibinfo {author} {\bibfnamefont {Y.}~\bibnamefont {Qi}}, \bibinfo {author} {\bibfnamefont {C.}~\bibnamefont {Shekhar}}, \bibinfo {author} {\bibfnamefont {Y.}~\bibnamefont {Sun}}, \bibinfo {author} {\bibfnamefont {Z.}~\bibnamefont {Wang}}, \emph {et~al.},\ }\bibfield  {title} {\bibinfo {title} {Axionic charge-density wave in the {W}eyl semimetal ({TaSe}$_4$) {2I}},\ }\href {https://doi.org/https://doi.org/10.1038/s41586-019-1630-4} {\bibfield  {journal} {\bibinfo  {journal} {Nature}\ }\textbf {\bibinfo {volume} {575}},\ \bibinfo {pages} {315} (\bibinfo {year} {2019})}\BibitemShut {NoStop}%
\bibitem [{\citenamefont {Hsieh}\ \emph {et~al.}(2008)\citenamefont {Hsieh}, \citenamefont {Qian}, \citenamefont {Wray}, \citenamefont {Xia}, \citenamefont {Hor}, \citenamefont {Cava},\ and\ \citenamefont {Hasan}}]{TI3}%
  \BibitemOpen
  \bibfield  {author} {\bibinfo {author} {\bibfnamefont {D.}~\bibnamefont {Hsieh}}, \bibinfo {author} {\bibfnamefont {D.}~\bibnamefont {Qian}}, \bibinfo {author} {\bibfnamefont {L.}~\bibnamefont {Wray}}, \bibinfo {author} {\bibfnamefont {Y.}~\bibnamefont {Xia}}, \bibinfo {author} {\bibfnamefont {Y.~S.}\ \bibnamefont {Hor}}, \bibinfo {author} {\bibfnamefont {R.~J.}\ \bibnamefont {Cava}},\ and\ \bibinfo {author} {\bibfnamefont {M.~Z.}\ \bibnamefont {Hasan}},\ }\bibfield  {title} {\bibinfo {title} {A topological {D}irac insulator in a quantum spin {H}all phase},\ }\href {https://doi.org/https://doi.org/10.1038/nature06843} {\bibfield  {journal} {\bibinfo  {journal} {Nature}\ }\textbf {\bibinfo {volume} {452}},\ \bibinfo {pages} {970} (\bibinfo {year} {2008})}\BibitemShut {NoStop}%
\bibitem [{\citenamefont {Hsieh}\ \emph {et~al.}(2009)\citenamefont {Hsieh}, \citenamefont {Xia}, \citenamefont {Wray}, \citenamefont {Qian}, \citenamefont {Pal}, \citenamefont {Dil}, \citenamefont {Osterwalder}, \citenamefont {Meier}, \citenamefont {Bihlmayer}, \citenamefont {Kane} \emph {et~al.}}]{TI4}%
  \BibitemOpen
  \bibfield  {author} {\bibinfo {author} {\bibfnamefont {D.}~\bibnamefont {Hsieh}}, \bibinfo {author} {\bibfnamefont {Y.}~\bibnamefont {Xia}}, \bibinfo {author} {\bibfnamefont {L.}~\bibnamefont {Wray}}, \bibinfo {author} {\bibfnamefont {D.}~\bibnamefont {Qian}}, \bibinfo {author} {\bibfnamefont {A.}~\bibnamefont {Pal}}, \bibinfo {author} {\bibfnamefont {J.}~\bibnamefont {Dil}}, \bibinfo {author} {\bibfnamefont {J.}~\bibnamefont {Osterwalder}}, \bibinfo {author} {\bibfnamefont {F.}~\bibnamefont {Meier}}, \bibinfo {author} {\bibfnamefont {G.}~\bibnamefont {Bihlmayer}}, \bibinfo {author} {\bibfnamefont {C.}~\bibnamefont {Kane}}, \emph {et~al.},\ }\bibfield  {title} {\bibinfo {title} {Observation of unconventional quantum spin textures in topological insulators},\ }\href {https://doi.org/https://doi.org/10.1126/science.1167733} {\bibfield  {journal} {\bibinfo  {journal} {Science}\ }\textbf {\bibinfo {volume} {323}},\ \bibinfo {pages} {919} (\bibinfo {year} {2009})}\BibitemShut {NoStop}%
\bibitem [{\citenamefont {Mudgal}\ \emph {et~al.}(2024)\citenamefont {Mudgal}, \citenamefont {Dutta}, \citenamefont {Meena}, \citenamefont {Yenugonda}, \citenamefont {Tiwari}, \citenamefont {Malik}, \citenamefont {Buck}, \citenamefont {Mahatha}, \citenamefont {Agarwal},\ and\ \citenamefont {Nayak}}]{MS813}%
  \BibitemOpen
  \bibfield  {author} {\bibinfo {author} {\bibfnamefont {M.}~\bibnamefont {Mudgal}}, \bibinfo {author} {\bibfnamefont {D.}~\bibnamefont {Dutta}}, \bibinfo {author} {\bibfnamefont {P.}~\bibnamefont {Meena}}, \bibinfo {author} {\bibfnamefont {V.}~\bibnamefont {Yenugonda}}, \bibinfo {author} {\bibfnamefont {V.~K.}\ \bibnamefont {Tiwari}}, \bibinfo {author} {\bibfnamefont {V.~K.}\ \bibnamefont {Malik}}, \bibinfo {author} {\bibfnamefont {J.}~\bibnamefont {Buck}}, \bibinfo {author} {\bibfnamefont {S.~K.}\ \bibnamefont {Mahatha}}, \bibinfo {author} {\bibfnamefont {A.}~\bibnamefont {Agarwal}},\ and\ \bibinfo {author} {\bibfnamefont {J.}~\bibnamefont {Nayak}},\ }\bibfield  {title} {\bibinfo {title} {{Magnetotransport and electronic structure of the axion insulator ${\mathrm{MnSb}}_{8}{\mathrm{Te}}_{13}$}},\ }\href {https://doi.org/10.1103/PhysRevB.110.045124} {\bibfield  {journal} {\bibinfo  {journal} {Phys. Rev. B}\ }\textbf {\bibinfo {volume} {110}},\ \bibinfo {pages} {045124} (\bibinfo {year} {2024})}\BibitemShut
  {NoStop}%
\bibitem [{\citenamefont {Meena}\ \emph {et~al.}(2025)\citenamefont {Meena}, \citenamefont {Mudgal}, \citenamefont {Bagga}, \citenamefont {Tiwari}, \citenamefont {Malik}, \citenamefont {Yadav},\ and\ \citenamefont {Nayak}}]{PhysRevMaterials.9.044202}%
  \BibitemOpen
  \bibfield  {author} {\bibinfo {author} {\bibfnamefont {P.}~\bibnamefont {Meena}}, \bibinfo {author} {\bibfnamefont {M.}~\bibnamefont {Mudgal}}, \bibinfo {author} {\bibfnamefont {S.}~\bibnamefont {Bagga}}, \bibinfo {author} {\bibfnamefont {V.~K.}\ \bibnamefont {Tiwari}}, \bibinfo {author} {\bibfnamefont {V.~K.}\ \bibnamefont {Malik}}, \bibinfo {author} {\bibfnamefont {C.~S.}\ \bibnamefont {Yadav}},\ and\ \bibinfo {author} {\bibfnamefont {J.}~\bibnamefont {Nayak}},\ }\bibfield  {title} {\bibinfo {title} {Investigation of weak antilocalization and $\ensuremath{\pi}$-berry phase by quantum oscillation in ferromagnetic {W}eyl semimetals},\ }\href {https://doi.org/10.1103/PhysRevMaterials.9.044202} {\bibfield  {journal} {\bibinfo  {journal} {Phys. Rev. Mater.}\ }\textbf {\bibinfo {volume} {9}},\ \bibinfo {pages} {044202} (\bibinfo {year} {2025})}\BibitemShut {NoStop}%
\bibitem [{\citenamefont {Rahn}\ \emph {et~al.}(2018)\citenamefont {Rahn}, \citenamefont {Soh}, \citenamefont {Francoual}, \citenamefont {Veiga}, \citenamefont {Strempfer}, \citenamefont {Mardegan}, \citenamefont {Yan}, \citenamefont {Guo}, \citenamefont {Shi},\ and\ \citenamefont {Boothroyd}}]{PhysRevB.97.214422}%
  \BibitemOpen
  \bibfield  {author} {\bibinfo {author} {\bibfnamefont {M.~C.}\ \bibnamefont {Rahn}}, \bibinfo {author} {\bibfnamefont {J.-R.}\ \bibnamefont {Soh}}, \bibinfo {author} {\bibfnamefont {S.}~\bibnamefont {Francoual}}, \bibinfo {author} {\bibfnamefont {L.~S.~I.}\ \bibnamefont {Veiga}}, \bibinfo {author} {\bibfnamefont {J.}~\bibnamefont {Strempfer}}, \bibinfo {author} {\bibfnamefont {J.}~\bibnamefont {Mardegan}}, \bibinfo {author} {\bibfnamefont {D.~Y.}\ \bibnamefont {Yan}}, \bibinfo {author} {\bibfnamefont {Y.~F.}\ \bibnamefont {Guo}}, \bibinfo {author} {\bibfnamefont {Y.~G.}\ \bibnamefont {Shi}},\ and\ \bibinfo {author} {\bibfnamefont {A.~T.}\ \bibnamefont {Boothroyd}},\ }\bibfield  {title} {\bibinfo {title} {Coupling of magnetic order and charge transport in the candidate {D}irac semimetal {EuCd}$_{2}${As}$_{2}$},\ }\href {https://doi.org/10.1103/PhysRevB.97.214422} {\bibfield  {journal} {\bibinfo  {journal} {Phys. Rev. B}\ }\textbf {\bibinfo {volume} {97}},\ \bibinfo {pages} {214422} (\bibinfo {year}
  {2018})}\BibitemShut {NoStop}%
\bibitem [{\citenamefont {Du}\ \emph {et~al.}(2022)\citenamefont {Du}, \citenamefont {Yang}, \citenamefont {Nie}, \citenamefont {Wu}, \citenamefont {Li}, \citenamefont {Luo}, \citenamefont {Chen}, \citenamefont {Su}, \citenamefont {Smidman}, \citenamefont {Shi}, \citenamefont {Cao}, \citenamefont {Steglich}, \citenamefont {Song},\ and\ \citenamefont {Yuan}}]{Du2022}%
  \BibitemOpen
  \bibfield  {author} {\bibinfo {author} {\bibfnamefont {F.}~\bibnamefont {Du}}, \bibinfo {author} {\bibfnamefont {L.}~\bibnamefont {Yang}}, \bibinfo {author} {\bibfnamefont {Z.}~\bibnamefont {Nie}}, \bibinfo {author} {\bibfnamefont {N.}~\bibnamefont {Wu}}, \bibinfo {author} {\bibfnamefont {Y.}~\bibnamefont {Li}}, \bibinfo {author} {\bibfnamefont {S.}~\bibnamefont {Luo}}, \bibinfo {author} {\bibfnamefont {Y.}~\bibnamefont {Chen}}, \bibinfo {author} {\bibfnamefont {D.}~\bibnamefont {Su}}, \bibinfo {author} {\bibfnamefont {M.}~\bibnamefont {Smidman}}, \bibinfo {author} {\bibfnamefont {Y.}~\bibnamefont {Shi}}, \bibinfo {author} {\bibfnamefont {C.}~\bibnamefont {Cao}}, \bibinfo {author} {\bibfnamefont {F.}~\bibnamefont {Steglich}}, \bibinfo {author} {\bibfnamefont {Y.}~\bibnamefont {Song}},\ and\ \bibinfo {author} {\bibfnamefont {H.}~\bibnamefont {Yuan}},\ }\bibfield  {title} {\bibinfo {title} {Consecutive topological phase transitions and colossal magnetoresistance in a magnetic topological semimetal},\ }\href
  {https://doi.org/10.1038/s41535-022-00468-0} {\bibfield  {journal} {\bibinfo  {journal} {npj Quantum Mater.}\ }\textbf {\bibinfo {volume} {7}},\ \bibinfo {pages} {65} (\bibinfo {year} {2022})}\BibitemShut {NoStop}%
\bibitem [{\citenamefont {Wang}\ \emph {et~al.}(2021)\citenamefont {Wang}, \citenamefont {Rogers}, \citenamefont {Yao}, \citenamefont {Nichols}, \citenamefont {Atay}, \citenamefont {Xu}, \citenamefont {Franklin}, \citenamefont {Sochnikov}, \citenamefont {Ryan}, \citenamefont {Haskel},\ and\ \citenamefont {Tafti}}]{Wang}%
  \BibitemOpen
  \bibfield  {author} {\bibinfo {author} {\bibfnamefont {Z.-C.}\ \bibnamefont {Wang}}, \bibinfo {author} {\bibfnamefont {J.~D.}\ \bibnamefont {Rogers}}, \bibinfo {author} {\bibfnamefont {X.}~\bibnamefont {Yao}}, \bibinfo {author} {\bibfnamefont {R.}~\bibnamefont {Nichols}}, \bibinfo {author} {\bibfnamefont {K.}~\bibnamefont {Atay}}, \bibinfo {author} {\bibfnamefont {B.}~\bibnamefont {Xu}}, \bibinfo {author} {\bibfnamefont {J.}~\bibnamefont {Franklin}}, \bibinfo {author} {\bibfnamefont {I.}~\bibnamefont {Sochnikov}}, \bibinfo {author} {\bibfnamefont {P.~J.}\ \bibnamefont {Ryan}}, \bibinfo {author} {\bibfnamefont {D.}~\bibnamefont {Haskel}},\ and\ \bibinfo {author} {\bibfnamefont {F.}~\bibnamefont {Tafti}},\ }\bibfield  {title} {\bibinfo {title} {Colossal magnetoresistance without mixed valence in a layered phosphide crystal},\ }\href {https://doi.org/https://doi.org/10.1002/adma.202005755} {\bibfield  {journal} {\bibinfo  {journal} {Adv. Mater.}\ }\textbf {\bibinfo {volume} {33}},\ \bibinfo {pages} {2005755}
  (\bibinfo {year} {2021})}\BibitemShut {NoStop}%
\bibitem [{\citenamefont {Zhang}\ \emph {et~al.}(2023)\citenamefont {Zhang}, \citenamefont {Du}, \citenamefont {Zheng}, \citenamefont {Luo}, \citenamefont {Wu}, \citenamefont {Zheng}, \citenamefont {Cui}, \citenamefont {Sun}, \citenamefont {Liu}, \citenamefont {Shen}, \citenamefont {Smidman}, \citenamefont {Song}, \citenamefont {Shi}, \citenamefont {Zhong}, \citenamefont {Cao}, \citenamefont {Yuan},\ and\ \citenamefont {Liu}}]{PhysRevB.108.L241115}%
  \BibitemOpen
  \bibfield  {author} {\bibinfo {author} {\bibfnamefont {H.}~\bibnamefont {Zhang}}, \bibinfo {author} {\bibfnamefont {F.}~\bibnamefont {Du}}, \bibinfo {author} {\bibfnamefont {X.}~\bibnamefont {Zheng}}, \bibinfo {author} {\bibfnamefont {S.}~\bibnamefont {Luo}}, \bibinfo {author} {\bibfnamefont {Y.}~\bibnamefont {Wu}}, \bibinfo {author} {\bibfnamefont {H.}~\bibnamefont {Zheng}}, \bibinfo {author} {\bibfnamefont {S.}~\bibnamefont {Cui}}, \bibinfo {author} {\bibfnamefont {Z.}~\bibnamefont {Sun}}, \bibinfo {author} {\bibfnamefont {Z.}~\bibnamefont {Liu}}, \bibinfo {author} {\bibfnamefont {D.}~\bibnamefont {Shen}}, \bibinfo {author} {\bibfnamefont {M.}~\bibnamefont {Smidman}}, \bibinfo {author} {\bibfnamefont {Y.}~\bibnamefont {Song}}, \bibinfo {author} {\bibfnamefont {M.}~\bibnamefont {Shi}}, \bibinfo {author} {\bibfnamefont {Z.}~\bibnamefont {Zhong}}, \bibinfo {author} {\bibfnamefont {C.}~\bibnamefont {Cao}}, \bibinfo {author} {\bibfnamefont {H.}~\bibnamefont {Yuan}},\ and\ \bibinfo {author} {\bibfnamefont
  {Y.}~\bibnamefont {Liu}},\ }\bibfield  {title} {\bibinfo {title} {Electronic band reconstruction across the insulator-metal transition in colossally magnetoresistive {EuCd}$_{2}${P}$_{2}$},\ }\href {https://doi.org/10.1103/PhysRevB.108.L241115} {\bibfield  {journal} {\bibinfo  {journal} {Phys. Rev. B}\ }\textbf {\bibinfo {volume} {108}},\ \bibinfo {pages} {L241115} (\bibinfo {year} {2023})}\BibitemShut {NoStop}%
\bibitem [{\citenamefont {Roychowdhury}\ \emph {et~al.}(2023)\citenamefont {Roychowdhury}, \citenamefont {Yao}, \citenamefont {Samanta}, \citenamefont {Bae}, \citenamefont {Chen}, \citenamefont {Ju}, \citenamefont {Raghavan}, \citenamefont {Kumar}, \citenamefont {Constantinou}, \citenamefont {Guin}, \citenamefont {Plumb}, \citenamefont {Romanelli}, \citenamefont {Borrmann}, \citenamefont {Vergniory}, \citenamefont {Strocov}, \citenamefont {Madhavan}, \citenamefont {Shekhar},\ and\ \citenamefont {Felser}}]{Roy}%
  \BibitemOpen
  \bibfield  {author} {\bibinfo {author} {\bibfnamefont {S.}~\bibnamefont {Roychowdhury}}, \bibinfo {author} {\bibfnamefont {M.}~\bibnamefont {Yao}}, \bibinfo {author} {\bibfnamefont {K.}~\bibnamefont {Samanta}}, \bibinfo {author} {\bibfnamefont {S.}~\bibnamefont {Bae}}, \bibinfo {author} {\bibfnamefont {D.}~\bibnamefont {Chen}}, \bibinfo {author} {\bibfnamefont {S.}~\bibnamefont {Ju}}, \bibinfo {author} {\bibfnamefont {A.}~\bibnamefont {Raghavan}}, \bibinfo {author} {\bibfnamefont {N.}~\bibnamefont {Kumar}}, \bibinfo {author} {\bibfnamefont {P.}~\bibnamefont {Constantinou}}, \bibinfo {author} {\bibfnamefont {S.~N.}\ \bibnamefont {Guin}}, \bibinfo {author} {\bibfnamefont {N.~C.}\ \bibnamefont {Plumb}}, \bibinfo {author} {\bibfnamefont {M.}~\bibnamefont {Romanelli}}, \bibinfo {author} {\bibfnamefont {H.}~\bibnamefont {Borrmann}}, \bibinfo {author} {\bibfnamefont {M.~G.}\ \bibnamefont {Vergniory}}, \bibinfo {author} {\bibfnamefont {V.~N.}\ \bibnamefont {Strocov}}, \bibinfo {author} {\bibfnamefont
  {V.}~\bibnamefont {Madhavan}}, \bibinfo {author} {\bibfnamefont {C.}~\bibnamefont {Shekhar}},\ and\ \bibinfo {author} {\bibfnamefont {C.}~\bibnamefont {Felser}},\ }\bibfield  {title} {\bibinfo {title} {Anomalous {H}all conductivity and {N}ernst effect of the ideal {W}eyl semimetallic ferromagnet {EuCd}$_{2}${P}$_{2}$},\ }\href {https://doi.org/https://doi.org/10.1002/advs.202207121} {\bibfield  {journal} {\bibinfo  {journal} {Adv. Sci.}\ }\textbf {\bibinfo {volume} {10}},\ \bibinfo {pages} {2207121} (\bibinfo {year} {2023})}\BibitemShut {NoStop}%
\bibitem [{\citenamefont {Xu}\ \emph {et~al.}(2021)\citenamefont {Xu}, \citenamefont {Das}, \citenamefont {Ma}, \citenamefont {Yi}, \citenamefont {Nie}, \citenamefont {Shi}, \citenamefont {Tiwari}, \citenamefont {Tsirkin}, \citenamefont {Neupert}, \citenamefont {Medarde}, \citenamefont {Shi}, \citenamefont {Chang},\ and\ \citenamefont {Shang}}]{PhysRevLett.126.076602}%
  \BibitemOpen
  \bibfield  {author} {\bibinfo {author} {\bibfnamefont {Y.}~\bibnamefont {Xu}}, \bibinfo {author} {\bibfnamefont {L.}~\bibnamefont {Das}}, \bibinfo {author} {\bibfnamefont {J.~Z.}\ \bibnamefont {Ma}}, \bibinfo {author} {\bibfnamefont {C.~J.}\ \bibnamefont {Yi}}, \bibinfo {author} {\bibfnamefont {S.~M.}\ \bibnamefont {Nie}}, \bibinfo {author} {\bibfnamefont {Y.~G.}\ \bibnamefont {Shi}}, \bibinfo {author} {\bibfnamefont {A.}~\bibnamefont {Tiwari}}, \bibinfo {author} {\bibfnamefont {S.~S.}\ \bibnamefont {Tsirkin}}, \bibinfo {author} {\bibfnamefont {T.}~\bibnamefont {Neupert}}, \bibinfo {author} {\bibfnamefont {M.}~\bibnamefont {Medarde}}, \bibinfo {author} {\bibfnamefont {M.}~\bibnamefont {Shi}}, \bibinfo {author} {\bibfnamefont {J.}~\bibnamefont {Chang}},\ and\ \bibinfo {author} {\bibfnamefont {T.}~\bibnamefont {Shang}},\ }\bibfield  {title} {\bibinfo {title} {Unconventional transverse transport above and below the magnetic transition temperature in {W}eyl semimetal {EuCd}$_{2}${P}$_{2}$},\ }\href
  {https://doi.org/10.1103/PhysRevLett.126.076602} {\bibfield  {journal} {\bibinfo  {journal} {Phys. Rev. Lett.}\ }\textbf {\bibinfo {volume} {126}},\ \bibinfo {pages} {076602} (\bibinfo {year} {2021})}\BibitemShut {NoStop}%
\bibitem [{\citenamefont {Zhang}\ \emph {et~al.}(2020)\citenamefont {Zhang}, \citenamefont {Deng}, \citenamefont {Zhang}, \citenamefont {Wang}, \citenamefont {Wang}, \citenamefont {Liu}, \citenamefont {Mei}, \citenamefont {Kumar}, \citenamefont {Schwier}, \citenamefont {Shimada}, \citenamefont {Chen},\ and\ \citenamefont {Shen}}]{PhysRevB.101.205126}%
  \BibitemOpen
  \bibfield  {author} {\bibinfo {author} {\bibfnamefont {Y.}~\bibnamefont {Zhang}}, \bibinfo {author} {\bibfnamefont {K.}~\bibnamefont {Deng}}, \bibinfo {author} {\bibfnamefont {X.}~\bibnamefont {Zhang}}, \bibinfo {author} {\bibfnamefont {M.}~\bibnamefont {Wang}}, \bibinfo {author} {\bibfnamefont {Y.}~\bibnamefont {Wang}}, \bibinfo {author} {\bibfnamefont {C.}~\bibnamefont {Liu}}, \bibinfo {author} {\bibfnamefont {J.-W.}\ \bibnamefont {Mei}}, \bibinfo {author} {\bibfnamefont {S.}~\bibnamefont {Kumar}}, \bibinfo {author} {\bibfnamefont {E.~F.}\ \bibnamefont {Schwier}}, \bibinfo {author} {\bibfnamefont {K.}~\bibnamefont {Shimada}}, \bibinfo {author} {\bibfnamefont {C.}~\bibnamefont {Chen}},\ and\ \bibinfo {author} {\bibfnamefont {B.}~\bibnamefont {Shen}},\ }\bibfield  {title} {\bibinfo {title} {In-plane antiferromagnetic moments and magnetic polaron in the axion topological insulator candidate {EuIn}$_{2}${As}$_{2}$},\ }\href {https://doi.org/10.1103/PhysRevB.101.205126} {\bibfield  {journal} {\bibinfo
  {journal} {Phys. Rev. B}\ }\textbf {\bibinfo {volume} {101}},\ \bibinfo {pages} {205126} (\bibinfo {year} {2020})}\BibitemShut {NoStop}%
\bibitem [{\citenamefont {Regmi}\ \emph {et~al.}(2020)\citenamefont {Regmi}, \citenamefont {Hosen}, \citenamefont {Ghosh}, \citenamefont {Singh}, \citenamefont {Dhakal}, \citenamefont {Sims}, \citenamefont {Wang}, \citenamefont {Kabir}, \citenamefont {Dimitri}, \citenamefont {Liu}, \citenamefont {Agarwal}, \citenamefont {Lin}, \citenamefont {Kaczorowski}, \citenamefont {Bansil},\ and\ \citenamefont {Neupane}}]{PhysRevB.102.165153}%
  \BibitemOpen
  \bibfield  {author} {\bibinfo {author} {\bibfnamefont {S.}~\bibnamefont {Regmi}}, \bibinfo {author} {\bibfnamefont {M.~M.}\ \bibnamefont {Hosen}}, \bibinfo {author} {\bibfnamefont {B.}~\bibnamefont {Ghosh}}, \bibinfo {author} {\bibfnamefont {B.}~\bibnamefont {Singh}}, \bibinfo {author} {\bibfnamefont {G.}~\bibnamefont {Dhakal}}, \bibinfo {author} {\bibfnamefont {C.}~\bibnamefont {Sims}}, \bibinfo {author} {\bibfnamefont {B.}~\bibnamefont {Wang}}, \bibinfo {author} {\bibfnamefont {F.}~\bibnamefont {Kabir}}, \bibinfo {author} {\bibfnamefont {K.}~\bibnamefont {Dimitri}}, \bibinfo {author} {\bibfnamefont {Y.}~\bibnamefont {Liu}}, \bibinfo {author} {\bibfnamefont {A.}~\bibnamefont {Agarwal}}, \bibinfo {author} {\bibfnamefont {H.}~\bibnamefont {Lin}}, \bibinfo {author} {\bibfnamefont {D.}~\bibnamefont {Kaczorowski}}, \bibinfo {author} {\bibfnamefont {A.}~\bibnamefont {Bansil}},\ and\ \bibinfo {author} {\bibfnamefont {M.}~\bibnamefont {Neupane}},\ }\bibfield  {title} {\bibinfo {title} {Temperature-dependent
  electronic structure in a higher-order topological insulator candidate {EuIn}$_{2}${As}$_{2}$},\ }\href {https://doi.org/10.1103/PhysRevB.102.165153} {\bibfield  {journal} {\bibinfo  {journal} {Phys. Rev. B}\ }\textbf {\bibinfo {volume} {102}},\ \bibinfo {pages} {165153} (\bibinfo {year} {2020})}\BibitemShut {NoStop}%
\bibitem [{\citenamefont {Ram}\ \emph {et~al.}(2024)\citenamefont {Ram}, \citenamefont {Singh}, \citenamefont {Banerjee}, \citenamefont {Sundaresan}, \citenamefont {Samal}, \citenamefont {Kanchana},\ and\ \citenamefont {Hossain}}]{ram2024magnetotransport}%
  \BibitemOpen
  \bibfield  {author} {\bibinfo {author} {\bibfnamefont {D.}~\bibnamefont {Ram}}, \bibinfo {author} {\bibfnamefont {J.}~\bibnamefont {Singh}}, \bibinfo {author} {\bibfnamefont {S.}~\bibnamefont {Banerjee}}, \bibinfo {author} {\bibfnamefont {A.}~\bibnamefont {Sundaresan}}, \bibinfo {author} {\bibfnamefont {D.}~\bibnamefont {Samal}}, \bibinfo {author} {\bibfnamefont {V.}~\bibnamefont {Kanchana}},\ and\ \bibinfo {author} {\bibfnamefont {Z.}~\bibnamefont {Hossain}},\ }\bibfield  {title} {\bibinfo {title} {Magnetotransport and electronic structure of {EuAuSb}: A candidate antiferromagnetic {D}irac semimetal},\ }\href {https://doi.org/https://doi.org/10.1103/PhysRevB.109.155152} {\bibfield  {journal} {\bibinfo  {journal} {Phys. Rev. B}\ }\textbf {\bibinfo {volume} {109}},\ \bibinfo {pages} {155152} (\bibinfo {year} {2024})}\BibitemShut {NoStop}%
\bibitem [{\citenamefont {Sears}\ \emph {et~al.}(2025)\citenamefont {Sears}, \citenamefont {Yao}, \citenamefont {Hu}, \citenamefont {Tian}, \citenamefont {Aryal}, \citenamefont {Yin}, \citenamefont {Tsvelik}, \citenamefont {Zaliznyak}, \citenamefont {Li},\ and\ \citenamefont {Tranquada}}]{sears2025euausb}%
  \BibitemOpen
  \bibfield  {author} {\bibinfo {author} {\bibfnamefont {J.}~\bibnamefont {Sears}}, \bibinfo {author} {\bibfnamefont {J.}~\bibnamefont {Yao}}, \bibinfo {author} {\bibfnamefont {Z.}~\bibnamefont {Hu}}, \bibinfo {author} {\bibfnamefont {W.}~\bibnamefont {Tian}}, \bibinfo {author} {\bibfnamefont {N.}~\bibnamefont {Aryal}}, \bibinfo {author} {\bibfnamefont {W.}~\bibnamefont {Yin}}, \bibinfo {author} {\bibfnamefont {A.}~\bibnamefont {Tsvelik}}, \bibinfo {author} {\bibfnamefont {I.~A.}\ \bibnamefont {Zaliznyak}}, \bibinfo {author} {\bibfnamefont {Q.}~\bibnamefont {Li}},\ and\ \bibinfo {author} {\bibfnamefont {J.~M.}\ \bibnamefont {Tranquada}},\ }\bibfield  {title} {\bibinfo {title} {{EuAuSb}: An odd-parity helical variation of altermagnetism},\ }\href {https://doi.org/https://doi.org/10.1103/k376-4cxw} {\bibfield  {journal} {\bibinfo  {journal} {Phys. Rev. B}\ }\textbf {\bibinfo {volume} {112}},\ \bibinfo {pages} {094455} (\bibinfo {year} {2025})}\BibitemShut {NoStop}%
\bibitem [{\citenamefont {Zhu}\ \emph {et~al.}(2020)\citenamefont {Zhu}, \citenamefont {Wang}, \citenamefont {Meven}, \citenamefont {Song}, \citenamefont {Mueller}, \citenamefont {Yi}, \citenamefont {Ji}, \citenamefont {Shi}, \citenamefont {Ma}, \citenamefont {Schmalzl} \emph {et~al.}}]{zhu2020magnetic}%
  \BibitemOpen
  \bibfield  {author} {\bibinfo {author} {\bibfnamefont {F.}~\bibnamefont {Zhu}}, \bibinfo {author} {\bibfnamefont {X.}~\bibnamefont {Wang}}, \bibinfo {author} {\bibfnamefont {M.}~\bibnamefont {Meven}}, \bibinfo {author} {\bibfnamefont {J.}~\bibnamefont {Song}}, \bibinfo {author} {\bibfnamefont {T.}~\bibnamefont {Mueller}}, \bibinfo {author} {\bibfnamefont {C.}~\bibnamefont {Yi}}, \bibinfo {author} {\bibfnamefont {W.}~\bibnamefont {Ji}}, \bibinfo {author} {\bibfnamefont {Y.}~\bibnamefont {Shi}}, \bibinfo {author} {\bibfnamefont {J.}~\bibnamefont {Ma}}, \bibinfo {author} {\bibfnamefont {K.}~\bibnamefont {Schmalzl}}, \emph {et~al.},\ }\bibfield  {title} {\bibinfo {title} {Magnetic structures, spin-flop transition, and coupling of {Eu} and {Mn} magnetism in the {D}irac semimetal {EuMnBi}$_2$},\ }\href {https://doi.org/https://doi.org/10.1103/PhysRevResearch.2.043100} {\bibfield  {journal} {\bibinfo  {journal} {Phys. Rev. Res.}\ }\textbf {\bibinfo {volume} {2}},\ \bibinfo {pages} {043100} (\bibinfo {year}
  {2020})}\BibitemShut {NoStop}%
\bibitem [{\citenamefont {Laha}\ \emph {et~al.}(2021)\citenamefont {Laha}, \citenamefont {Singha}, \citenamefont {Mardanya}, \citenamefont {Singh}, \citenamefont {Agarwal}, \citenamefont {Mandal},\ and\ \citenamefont {Hossain}}]{PhysRevB.103.L241112}%
  \BibitemOpen
  \bibfield  {author} {\bibinfo {author} {\bibfnamefont {A.}~\bibnamefont {Laha}}, \bibinfo {author} {\bibfnamefont {R.}~\bibnamefont {Singha}}, \bibinfo {author} {\bibfnamefont {S.}~\bibnamefont {Mardanya}}, \bibinfo {author} {\bibfnamefont {B.}~\bibnamefont {Singh}}, \bibinfo {author} {\bibfnamefont {A.}~\bibnamefont {Agarwal}}, \bibinfo {author} {\bibfnamefont {P.}~\bibnamefont {Mandal}},\ and\ \bibinfo {author} {\bibfnamefont {Z.}~\bibnamefont {Hossain}},\ }\bibfield  {title} {\bibinfo {title} {Topological {H}all effect in the antiferromagnetic {D}irac semimetal {EuAgAs}},\ }\href {https://doi.org/10.1103/PhysRevB.103.L241112} {\bibfield  {journal} {\bibinfo  {journal} {Phys. Rev. B}\ }\textbf {\bibinfo {volume} {103}},\ \bibinfo {pages} {L241112} (\bibinfo {year} {2021})}\BibitemShut {NoStop}%
\bibitem [{\citenamefont {Jin}\ \emph {et~al.}(2021)\citenamefont {Jin}, \citenamefont {Zeng}, \citenamefont {Feng}, \citenamefont {Du}, \citenamefont {Wu}, \citenamefont {Sheng}, \citenamefont {Yu}, \citenamefont {Zhu},\ and\ \citenamefont {Yang}}]{PhysRevB.104.165424}%
  \BibitemOpen
  \bibfield  {author} {\bibinfo {author} {\bibfnamefont {Y.}~\bibnamefont {Jin}}, \bibinfo {author} {\bibfnamefont {X.-T.}\ \bibnamefont {Zeng}}, \bibinfo {author} {\bibfnamefont {X.}~\bibnamefont {Feng}}, \bibinfo {author} {\bibfnamefont {X.}~\bibnamefont {Du}}, \bibinfo {author} {\bibfnamefont {W.}~\bibnamefont {Wu}}, \bibinfo {author} {\bibfnamefont {X.-L.}\ \bibnamefont {Sheng}}, \bibinfo {author} {\bibfnamefont {Z.-M.}\ \bibnamefont {Yu}}, \bibinfo {author} {\bibfnamefont {Z.}~\bibnamefont {Zhu}},\ and\ \bibinfo {author} {\bibfnamefont {S.~A.}\ \bibnamefont {Yang}},\ }\bibfield  {title} {\bibinfo {title} {Multiple magnetism-controlled topological states in {EuAgAs}},\ }\href {https://doi.org/10.1103/PhysRevB.104.165424} {\bibfield  {journal} {\bibinfo  {journal} {Phys. Rev. B}\ }\textbf {\bibinfo {volume} {104}},\ \bibinfo {pages} {165424} (\bibinfo {year} {2021})}\BibitemShut {NoStop}%
\bibitem [{\citenamefont {Liu}\ \emph {et~al.}(2025)\citenamefont {Liu}, \citenamefont {Zhang}, \citenamefont {Wu}, \citenamefont {Jin}, \citenamefont {Zhu}, \citenamefont {Song}, \citenamefont {Cui}, \citenamefont {Sun}, \citenamefont {Wang}, \citenamefont {Chen}, \citenamefont {He}, \citenamefont {Liu}, \citenamefont {Duan}, \citenamefont {Oppeneer},\ and\ \citenamefont {Meng}}]{PhysRevB.111.L121113}%
  \BibitemOpen
  \bibfield  {author} {\bibinfo {author} {\bibfnamefont {H.}~\bibnamefont {Liu}}, \bibinfo {author} {\bibfnamefont {C.}~\bibnamefont {Zhang}}, \bibinfo {author} {\bibfnamefont {Q.-Y.}\ \bibnamefont {Wu}}, \bibinfo {author} {\bibfnamefont {Y.}~\bibnamefont {Jin}}, \bibinfo {author} {\bibfnamefont {Z.}~\bibnamefont {Zhu}}, \bibinfo {author} {\bibfnamefont {J.-J.}\ \bibnamefont {Song}}, \bibinfo {author} {\bibfnamefont {S.-T.}\ \bibnamefont {Cui}}, \bibinfo {author} {\bibfnamefont {Z.}~\bibnamefont {Sun}}, \bibinfo {author} {\bibfnamefont {H.}~\bibnamefont {Wang}}, \bibinfo {author} {\bibfnamefont {B.}~\bibnamefont {Chen}}, \bibinfo {author} {\bibfnamefont {J.}~\bibnamefont {He}}, \bibinfo {author} {\bibfnamefont {H.-Y.}\ \bibnamefont {Liu}}, \bibinfo {author} {\bibfnamefont {Y.-X.}\ \bibnamefont {Duan}}, \bibinfo {author} {\bibfnamefont {P.~M.}\ \bibnamefont {Oppeneer}},\ and\ \bibinfo {author} {\bibfnamefont {J.-Q.}\ \bibnamefont {Meng}},\ }\bibfield  {title} {\bibinfo {title} {Ultrafast photoinduced phase
  transition in the antiferromagnetic {D}irac semimetal {EuAgAs}},\ }\href {https://doi.org/10.1103/PhysRevB.111.L121113} {\bibfield  {journal} {\bibinfo  {journal} {Phys. Rev. B}\ }\textbf {\bibinfo {volume} {111}},\ \bibinfo {pages} {L121113} (\bibinfo {year} {2025})}\BibitemShut {NoStop}%
\bibitem [{\citenamefont {Gazzah}\ \emph {et~al.}(2026)\citenamefont {Gazzah}, \citenamefont {Kaplan}, \citenamefont {Morgan}, \citenamefont {Regmi}, \citenamefont {Jamaluddin}, \citenamefont {Cao}, \citenamefont {Mazin},\ and\ \citenamefont {Ghimire}}]{gazzah2026}%
  \BibitemOpen
  \bibfield  {author} {\bibinfo {author} {\bibfnamefont {M.~E.}\ \bibnamefont {Gazzah}}, \bibinfo {author} {\bibfnamefont {D.}~\bibnamefont {Kaplan}}, \bibinfo {author} {\bibfnamefont {Z.}~\bibnamefont {Morgan}}, \bibinfo {author} {\bibfnamefont {A.~N.~R.}\ \bibnamefont {Regmi}}, \bibinfo {author} {\bibfnamefont {S.}~\bibnamefont {Jamaluddin}}, \bibinfo {author} {\bibfnamefont {H.}~\bibnamefont {Cao}}, \bibinfo {author} {\bibfnamefont {I.~I.}\ \bibnamefont {Mazin}},\ and\ \bibinfo {author} {\bibfnamefont {N.~J.}\ \bibnamefont {Ghimire}},\ }\href@noop {} {\bibinfo {title} {Near-degenerate competing magnetic orders in {EuAgAs}: a tunable route to altermagnetism}} (\bibinfo {year} {2026}),\ \bibinfo {note} {\href{https://arxiv.org/abs/2605.16242}{arXiv:2605.16242}}\BibitemShut {NoStop}%
\bibitem [{\citenamefont {Kresse}\ and\ \citenamefont {Furthm\"uller}(1996)}]{PhysRevB.54.11169}%
  \BibitemOpen
  \bibfield  {author} {\bibinfo {author} {\bibfnamefont {G.}~\bibnamefont {Kresse}}\ and\ \bibinfo {author} {\bibfnamefont {J.}~\bibnamefont {Furthm\"uller}},\ }\bibfield  {title} {\bibinfo {title} {Efficient iterative schemes for ab initio total-energy calculations using a plane-wave basis set},\ }\href {https://doi.org/10.1103/PhysRevB.54.11169} {\bibfield  {journal} {\bibinfo  {journal} {Phys. Rev. B}\ }\textbf {\bibinfo {volume} {54}},\ \bibinfo {pages} {11169} (\bibinfo {year} {1996})}\BibitemShut {NoStop}%
\bibitem [{\citenamefont {Kresse}\ and\ \citenamefont {Joubert}(1999)}]{PhysRevB.59.1758}%
  \BibitemOpen
  \bibfield  {author} {\bibinfo {author} {\bibfnamefont {G.}~\bibnamefont {Kresse}}\ and\ \bibinfo {author} {\bibfnamefont {D.}~\bibnamefont {Joubert}},\ }\bibfield  {title} {\bibinfo {title} {From ultrasoft pseudopotentials to the projector augmented-wave method},\ }\href {https://doi.org/10.1103/PhysRevB.59.1758} {\bibfield  {journal} {\bibinfo  {journal} {Phys. Rev. B}\ }\textbf {\bibinfo {volume} {59}},\ \bibinfo {pages} {1758} (\bibinfo {year} {1999})}\BibitemShut {NoStop}%
\bibitem [{\citenamefont {Perdew}\ \emph {et~al.}(1996)\citenamefont {Perdew}, \citenamefont {Burke},\ and\ \citenamefont {Ernzerhof}}]{PhysRevLett.77.3865}%
  \BibitemOpen
  \bibfield  {author} {\bibinfo {author} {\bibfnamefont {J.~P.}\ \bibnamefont {Perdew}}, \bibinfo {author} {\bibfnamefont {K.}~\bibnamefont {Burke}},\ and\ \bibinfo {author} {\bibfnamefont {M.}~\bibnamefont {Ernzerhof}},\ }\bibfield  {title} {\bibinfo {title} {Generalized gradient approximation made simple},\ }\href {https://doi.org/10.1103/PhysRevLett.77.3865} {\bibfield  {journal} {\bibinfo  {journal} {Phys. Rev. Lett.}\ }\textbf {\bibinfo {volume} {77}},\ \bibinfo {pages} {3865} (\bibinfo {year} {1996})}\BibitemShut {NoStop}%
\bibitem [{\citenamefont {Mostofi}\ \emph {et~al.}(2008)\citenamefont {Mostofi}, \citenamefont {Yates}, \citenamefont {Lee}, \citenamefont {Souza}, \citenamefont {Vanderbilt},\ and\ \citenamefont {Marzari}}]{MOSTOFI2008685}%
  \BibitemOpen
  \bibfield  {author} {\bibinfo {author} {\bibfnamefont {A.~A.}\ \bibnamefont {Mostofi}}, \bibinfo {author} {\bibfnamefont {J.~R.}\ \bibnamefont {Yates}}, \bibinfo {author} {\bibfnamefont {Y.-S.}\ \bibnamefont {Lee}}, \bibinfo {author} {\bibfnamefont {I.}~\bibnamefont {Souza}}, \bibinfo {author} {\bibfnamefont {D.}~\bibnamefont {Vanderbilt}},\ and\ \bibinfo {author} {\bibfnamefont {N.}~\bibnamefont {Marzari}},\ }\bibfield  {title} {\bibinfo {title} {wannier90: A tool for obtaining maximally-localised {W}annier functions},\ }\href {https://doi.org/https://doi.org/10.1016/j.cpc.2007.11.016} {\bibfield  {journal} {\bibinfo  {journal} {Comput. Phys. Commun.}\ }\textbf {\bibinfo {volume} {178}},\ \bibinfo {pages} {685} (\bibinfo {year} {2008})}\BibitemShut {NoStop}%
\bibitem [{\citenamefont {Wu}\ \emph {et~al.}(2018)\citenamefont {Wu}, \citenamefont {Zhang}, \citenamefont {Song}, \citenamefont {Troyer},\ and\ \citenamefont {Soluyanov}}]{WU2018405}%
  \BibitemOpen
  \bibfield  {author} {\bibinfo {author} {\bibfnamefont {Q.}~\bibnamefont {Wu}}, \bibinfo {author} {\bibfnamefont {S.}~\bibnamefont {Zhang}}, \bibinfo {author} {\bibfnamefont {H.-F.}\ \bibnamefont {Song}}, \bibinfo {author} {\bibfnamefont {M.}~\bibnamefont {Troyer}},\ and\ \bibinfo {author} {\bibfnamefont {A.~A.}\ \bibnamefont {Soluyanov}},\ }\bibfield  {title} {\bibinfo {title} {Wanniertools: An open-source software package for novel topological materials},\ }\href {https://doi.org/https://doi.org/10.1016/j.cpc.2017.09.033} {\bibfield  {journal} {\bibinfo  {journal} {Comput. Phys. Commun.}\ }\textbf {\bibinfo {volume} {224}},\ \bibinfo {pages} {405} (\bibinfo {year} {2018})}\BibitemShut {NoStop}%
\bibitem [{\citenamefont {Yang}\ \emph {et~al.}(2015)\citenamefont {Yang}, \citenamefont {Morimoto},\ and\ \citenamefont {Furusaki}}]{Yang2015}%
  \BibitemOpen
  \bibfield  {author} {\bibinfo {author} {\bibfnamefont {B.-J.}\ \bibnamefont {Yang}}, \bibinfo {author} {\bibfnamefont {T.}~\bibnamefont {Morimoto}},\ and\ \bibinfo {author} {\bibfnamefont {A.}~\bibnamefont {Furusaki}},\ }\bibfield  {title} {\bibinfo {title} {Topological charges of three-dimensional {D}irac semimetals with rotation symmetry},\ }\href {https://doi.org/10.1103/PhysRevB.92.165120} {\bibfield  {journal} {\bibinfo  {journal} {Phy. Rev. B}\ }\textbf {\bibinfo {volume} {92}},\ \bibinfo {pages} {165120} (\bibinfo {year} {2015})}\BibitemShut {NoStop}%
\bibitem [{\citenamefont {Moser}(2017)}]{moser2017experimentalist}%
  \BibitemOpen
  \bibfield  {author} {\bibinfo {author} {\bibfnamefont {S.}~\bibnamefont {Moser}},\ }\bibfield  {title} {\bibinfo {title} {An experimentalist's guide to the matrix element in angle resolved photoemission},\ }\href {https://doi.org/https://doi.org/10.1016/j.elspec.2016.11.007} {\bibfield  {journal} {\bibinfo  {journal} {J. Electron Spectrosc. Relat. Phenom.}\ }\textbf {\bibinfo {volume} {214}},\ \bibinfo {pages} {29} (\bibinfo {year} {2017})}\BibitemShut {NoStop}%
\bibitem [{\citenamefont {Zhang}\ \emph {et~al.}(2022)\citenamefont {Zhang}, \citenamefont {Pincelli}, \citenamefont {Jozwiak}, \citenamefont {Kondo}, \citenamefont {Ernstorfer}, \citenamefont {Sato},\ and\ \citenamefont {Zhou}}]{zhang2022angle}%
  \BibitemOpen
  \bibfield  {author} {\bibinfo {author} {\bibfnamefont {H.}~\bibnamefont {Zhang}}, \bibinfo {author} {\bibfnamefont {T.}~\bibnamefont {Pincelli}}, \bibinfo {author} {\bibfnamefont {C.}~\bibnamefont {Jozwiak}}, \bibinfo {author} {\bibfnamefont {T.}~\bibnamefont {Kondo}}, \bibinfo {author} {\bibfnamefont {R.}~\bibnamefont {Ernstorfer}}, \bibinfo {author} {\bibfnamefont {T.}~\bibnamefont {Sato}},\ and\ \bibinfo {author} {\bibfnamefont {S.}~\bibnamefont {Zhou}},\ }\bibfield  {title} {\bibinfo {title} {Angle-resolved photoemission spectroscopy},\ }\href {https://doi.org/https://doi.org/10.1038/s43586-022-00133-7} {\bibfield  {journal} {\bibinfo  {journal} {Nat. Rev. Methods Primers}\ }\textbf {\bibinfo {volume} {2}},\ \bibinfo {pages} {54} (\bibinfo {year} {2022})}\BibitemShut {NoStop}%
\bibitem [{Sup()}]{Suppl}%
  \BibitemOpen
  \href@noop {} {}\bibinfo {note} {See Supplementary Information [URL will be inserted by publisher] for polarisation dependent measurement geometry, additional temperature dependent ARPES Fermi surfaces, Band dispersion, DFT calculated Fermi surfaces at different $k_z$ values, orbital contributions from different elements to the Fermi surface, and a discussion of the higher order vHs.}\BibitemShut {Stop}%
\bibitem [{\citenamefont {Hossain}\ \emph {et~al.}(2025)\citenamefont {Hossain}, \citenamefont {Islam}, \citenamefont {Cheng}, \citenamefont {Muhammad}, \citenamefont {Zhang}, \citenamefont {Guguchia}, \citenamefont {Krieger}, \citenamefont {Casas}, \citenamefont {Jiang}, \citenamefont {Litskevich} \emph {et~al.}}]{hossain2025superconductivity}%
  \BibitemOpen
  \bibfield  {author} {\bibinfo {author} {\bibfnamefont {M.~S.}\ \bibnamefont {Hossain}}, \bibinfo {author} {\bibfnamefont {R.}~\bibnamefont {Islam}}, \bibinfo {author} {\bibfnamefont {Z.-J.}\ \bibnamefont {Cheng}}, \bibinfo {author} {\bibfnamefont {Z.}~\bibnamefont {Muhammad}}, \bibinfo {author} {\bibfnamefont {Q.}~\bibnamefont {Zhang}}, \bibinfo {author} {\bibfnamefont {Z.}~\bibnamefont {Guguchia}}, \bibinfo {author} {\bibfnamefont {J.~A.}\ \bibnamefont {Krieger}}, \bibinfo {author} {\bibfnamefont {B.}~\bibnamefont {Casas}}, \bibinfo {author} {\bibfnamefont {Y.-X.}\ \bibnamefont {Jiang}}, \bibinfo {author} {\bibfnamefont {M.}~\bibnamefont {Litskevich}}, \emph {et~al.},\ }\bibfield  {title} {\bibinfo {title} {Superconductivity and a van {H}ove singularity confined to the surface of a topological semimetal},\ }\href {https://doi.org/https://doi.org/10.1038/s41467-025-58024-w} {\bibfield  {journal} {\bibinfo  {journal} {Nat. commun.}\ }\textbf {\bibinfo {volume} {16}},\ \bibinfo {pages} {3998} (\bibinfo {year}
  {2025})}\BibitemShut {NoStop}%
\bibitem [{\citenamefont {Hu}\ \emph {et~al.}(2022)\citenamefont {Hu}, \citenamefont {Wu}, \citenamefont {Ortiz}, \citenamefont {Ju}, \citenamefont {Han}, \citenamefont {Ma}, \citenamefont {Plumb}, \citenamefont {Radovic}, \citenamefont {Thomale}, \citenamefont {Wilson} \emph {et~al.}}]{hu2022rich}%
  \BibitemOpen
  \bibfield  {author} {\bibinfo {author} {\bibfnamefont {Y.}~\bibnamefont {Hu}}, \bibinfo {author} {\bibfnamefont {X.}~\bibnamefont {Wu}}, \bibinfo {author} {\bibfnamefont {B.~R.}\ \bibnamefont {Ortiz}}, \bibinfo {author} {\bibfnamefont {S.}~\bibnamefont {Ju}}, \bibinfo {author} {\bibfnamefont {X.}~\bibnamefont {Han}}, \bibinfo {author} {\bibfnamefont {J.}~\bibnamefont {Ma}}, \bibinfo {author} {\bibfnamefont {N.~C.}\ \bibnamefont {Plumb}}, \bibinfo {author} {\bibfnamefont {M.}~\bibnamefont {Radovic}}, \bibinfo {author} {\bibfnamefont {R.}~\bibnamefont {Thomale}}, \bibinfo {author} {\bibfnamefont {S.~D.}\ \bibnamefont {Wilson}}, \emph {et~al.},\ }\bibfield  {title} {\bibinfo {title} {Rich nature of van {H}ove singularities in {K}agome superconductor {CsV}$_3${Sb}$_5$},\ }\href {https://doi.org/https://doi.org/10.1038/s41467-022-29828-x} {\bibfield  {journal} {\bibinfo  {journal} {Nat. Commun.}\ }\textbf {\bibinfo {volume} {13}},\ \bibinfo {pages} {2220} (\bibinfo {year} {2022})}\BibitemShut {NoStop}%
\bibitem [{\citenamefont {Louat}\ \emph {et~al.}(2019)\citenamefont {Louat}, \citenamefont {Lenz}, \citenamefont {Biermann}, \citenamefont {Martins}, \citenamefont {Bertran}, \citenamefont {Le~F\`evre}, \citenamefont {Rault}, \citenamefont {Bert},\ and\ \citenamefont {Brouet}}]{PhysRevB.100.205135}%
  \BibitemOpen
  \bibfield  {author} {\bibinfo {author} {\bibfnamefont {A.}~\bibnamefont {Louat}}, \bibinfo {author} {\bibfnamefont {B.}~\bibnamefont {Lenz}}, \bibinfo {author} {\bibfnamefont {S.}~\bibnamefont {Biermann}}, \bibinfo {author} {\bibfnamefont {C.}~\bibnamefont {Martins}}, \bibinfo {author} {\bibfnamefont {F.~m.~c.}\ \bibnamefont {Bertran}}, \bibinfo {author} {\bibfnamefont {P.}~\bibnamefont {Le~F\`evre}}, \bibinfo {author} {\bibfnamefont {J.~E.}\ \bibnamefont {Rault}}, \bibinfo {author} {\bibfnamefont {F.}~\bibnamefont {Bert}},\ and\ \bibinfo {author} {\bibfnamefont {V.}~\bibnamefont {Brouet}},\ }\bibfield  {title} {\bibinfo {title} {Arpes study of orbital character, symmetry breaking, and pseudogaps in doped and pure {Sr}$_2${IrO}$_4$},\ }\href {https://doi.org/10.1103/PhysRevB.100.205135} {\bibfield  {journal} {\bibinfo  {journal} {Phys. Rev. B}\ }\textbf {\bibinfo {volume} {100}},\ \bibinfo {pages} {205135} (\bibinfo {year} {2019})}\BibitemShut {NoStop}%
\bibitem [{\citenamefont {Soh}\ \emph {et~al.}(2026)\citenamefont {Soh}, \citenamefont {Zhu}, \citenamefont {Withers}, \citenamefont {Rodríguez-Velamazán}, \citenamefont {Kim}, \citenamefont {Fabelo}, \citenamefont {Stunault}, \citenamefont {Yevtushynsky}, \citenamefont {Prabhakaran}, \citenamefont {Yang},\ and\ \citenamefont {Boothroyd}}]{soh2026}%
  \BibitemOpen
  \bibfield  {author} {\bibinfo {author} {\bibfnamefont {J.-R.}\ \bibnamefont {Soh}}, \bibinfo {author} {\bibfnamefont {Z.}~\bibnamefont {Zhu}}, \bibinfo {author} {\bibfnamefont {L.}~\bibnamefont {Withers}}, \bibinfo {author} {\bibfnamefont {J.~A.}\ \bibnamefont {Rodríguez-Velamazán}}, \bibinfo {author} {\bibfnamefont {T.~K.}\ \bibnamefont {Kim}}, \bibinfo {author} {\bibfnamefont {O.}~\bibnamefont {Fabelo}}, \bibinfo {author} {\bibfnamefont {A.}~\bibnamefont {Stunault}}, \bibinfo {author} {\bibfnamefont {D.}~\bibnamefont {Yevtushynsky}}, \bibinfo {author} {\bibfnamefont {D.}~\bibnamefont {Prabhakaran}}, \bibinfo {author} {\bibfnamefont {S.~A.}\ \bibnamefont {Yang}},\ and\ \bibinfo {author} {\bibfnamefont {A.~T.}\ \bibnamefont {Boothroyd}},\ }\href@noop {} {\bibinfo {title} {Higher-order {W}eyl nodes driven by helical magnetic order in {EuAgAs}}} (\bibinfo {year} {2026}),\ \bibinfo {note} {\href{https://arxiv.org/abs/2605.19347}{arXiv:2605.19347}}\BibitemShut {NoStop}%
\end{thebibliography}

%

\end{document}